\documentclass[10pt,journal,twocolumn]{IEEEtran}
\IEEEoverridecommandlockouts
\usepackage{epsfig,cite,bm,algorithm,algorithmic,epstopdf,amsmath,subfigure,multirow,amssymb,amsfonts,amsthm,xcolor,caption,graphicx,textcomp,xcolor, subfigure, makecell, bbding}
\usepackage[inline, shortlabels]{enumitem}
\usepackage[flushleft]{threeparttable}
\setlist{itemjoin ={,\enspace},itemjoin* = { and\enspace}}

\usepackage[T1]{fontenc}
\usepackage{stfloats}
\allowdisplaybreaks[4]

\newcommand{\blue}[1]{\textcolor{blue}{#1}}
\allowdisplaybreaks[4]
\usepackage{booktabs}

\begin{document}

\title{Hybrid Beamforming Design for RSMA-enabled Near-Field Integrated Sensing and Communications}

\author{Jiasi Zhou, Chintha Tellambura,~\IEEEmembership{Fellow,~IEEE}, and  Geoffrey Ye Li, \IEEEmembership{Fellow,~IEEE} 

\thanks{Jiasi Zhou is with the School of Medical Information and Engineering, Xuzhou Medical University, Xuzhou, 221004, China, (email: jiasi\_zhou@xzhmu.edu.cn). (\emph{Corresponding author: Jiasi Zhou}).}
\thanks{ Chintha Tellambura is with the Department of Electrical and Computer Engineering, University of Alberta, Edmonton, AB, T6G 2R3, Canada (email: ct4@ualberta.ca).} 
\thanks{Geoffrey Ye Li is with the School of Electrical and Electronic Engineering, Imperial College London, London SW7 2AZ, UK (e-mail: geoffrey.li@imperial.ac.uk).}
\thanks{This work was supported by the Talented Scientific Research Foundation of Xuzhou Medical University (D2022027).}}
\maketitle
	
\begin{abstract}
Integrated sensing and communication (ISAC) networks leverage extremely large antenna arrays and high frequencies. This inevitably extends the Rayleigh distance, making near-field (NF) spherical wave propagation dominant. This unlocks numerous spatial degrees of freedom, raising the challenge of optimizing them for communication and sensing tradeoffs. To this end, we propose a rate-splitting multiple access (RSMA)-based NF-ISAC transmit scheme utilizing hybrid analog-digital antennas. RSMA enhances interference management, while a variable number of dedicated sensing beams adds beamforming flexibility. The objective is to maximize the minimum communication rate while ensuring multi-target sensing performance by jointly optimizing receive filters, analog and digital beamformers, common rate allocation, and the sensing beam count. To address uncertainty in sensing beam allocation, a rank-zero solution reconstruction method demonstrates that dedicated sensing beams are unnecessary for NF multi-target detection. A penalty dual decomposition (PDD)-based double-loop algorithm is introduced, employing weighted minimum mean-squared error (WMMSE) and quadratic transforms to reformulate communication and sensing rates. Simulations reveal that the proposed scheme: 1) achieves performance comparable to fully digital beamforming with fewer RF chains, (2) maintains NF multi-target detection without compromising communication rates, and 3) significantly outperforms conventional multiple access schemes and far-field ISAC systems.
\end{abstract} 


\section{Introduction}
Next-generation wireless networks will revolutionize integrated sensing and communication (ISAC) capabilities to support autonomous driving, indoor positioning, and smart infrastructure\cite{10663521}. ISAC facilitates the dual use of radio signals and wireless infrastructure, which is particularly relevant for these applications \cite{10663521,10559261}. However, the coexistence of dual functions creates a complex wireless environment, making effective interference management critical\cite{10522473}. Common approaches include space division multiple access (SDMA) and non-orthogonal multiple access (NOMA). SDMA treats interference as noise, while NOMA decodes all stronger interference. Both lack flexibility and fail to provide precise interference management, limiting network performance\cite{mao2018rate,10458958,10798456}. As a remedy, rate-splitting multiple access (RSMA) offers a more robust and adaptable solution, allowing receivers to partially decode interference while tolerating residual. By adjusting the interference decoding percentage, RSMA generalizes SDMA and NOMA as special cases \cite{mao2018rate,9831440}. 

Many works have demonstrated that RSMA delivers higher spectral efficiency and fairness than SDMA and NOMA in ISAC\cite{10251151,10382465,10380513,10679658}. However, future ISAC networks are advancing towards extremely large-scale antenna arrays (ELAA) and high-frequency bands to enhance communication capacity and sensing resolution\cite{10220205}. These technological advancements inevitably expand the Rayleigh distance, typically reaching several tens or even hundreds of meters.  As a result, the nature of electromagnetic (EM) propagation fundamentally shifts \textemdash  from far-field (FF) plane-wave characteristics to near-field (NF) spherical-wave behavior\cite{10559261}. Unlike plane-wave propagation, spherical waves introduce an additional distance dimension, incorporating both direction and distance information\cite{10559261}. This characteristic enables NF beamforming to focus energy on specific points, but several fundamental questions in NF-ISAC arise naturally, summarized as follows.
\begin{enumerate}
\item \textbf{Can RSMA enhance network performance in NF-ISAC?} Spherical-wave channels can concentrate radiated energy designated on spatial positions rather than merely directed along specific directions\cite{10135096,10559261}. Capitalizing on this characteristic, spherical waves can leverage richer spatial information to distinguish users, achieving precise signal enhancement and interference management. These capabilities suggest that NF-ISAC may potentially obviate the need for anti-interference schemes. However, the accuracy of this hypothesis has yet to be confirmed.

\item \textbf{ How many dedicated sensing beams are needed for near-field target detection?} ISAC inherently faces a tradeoff between communication and sensing tasks. Optimizing this tradeoff requires efficiently allocating spatial beams generated by the multi-antenna base station. The simulation results in\cite{9531484} show that the RSMA-enabled FF-ISAC does not require dedicated sensing beams. However, NF-ISAC utilizes spherical-wave-based beams, enhancing communication by focusing power on user positions \cite{10559261}. Such focused beams may not be well-suited for supporting NF multi-target detection, as their energy concentration limits broader sensing coverage. Consequently, additional dedicated sensing beams may be crucial to achieving optimal sensing performance in RSMA-enabled NF-ISAC systems. Nevertheless, there is currently no rigorous theoretical proof supporting this approach, and the exact number of required sensing beams remains unknown.

\item \textbf{How to alleviate hardware complexity?} Another significant challenge is the substantial radio frequency (RF) chain deployment. An RF chain comprises a power amplifier, a digital filter, a digital-to-analog converter (DAC), and a mixer. DACs dominate the total power consumption, and the RF chain is expensive\cite{10558818}. The NF-ISAC typically occurs in high-frequency and ELAA scenarios. Moreover, the fully digital beamforming requires each antenna to be connected to a dedicated RF chain\cite{10559261}. This results in substantially high design complexity and energy cost\cite{2024Towards}, so the fully digital beamforming becomes prohibitively expensive and power-thirsty. As a result, this calls for hybrid beamforming architectures.
\end{enumerate}

Answering these three issues can provide valuable insights for NF-ISAC  beamforming architecture selection and interference management. However, to our knowledge, these problems remain unexplored, motivating our work.

\subsection{Related works}
Prior relevant contributions can be divided into four categories, namely, SDMA/NOMA-enabled FF-ISAC\cite{10251151,10382465,10328645,10571110,10464353,10679658,10159012,9668964,10542219}, RSMA-enabled FF-ISAC\cite{9531484,10486996,10522473,10032141,10287099,10614103}, SDMA-enabled NF-ISAC\cite{10419546,10520715,10135096,10579914,gavras2024simultaneous}, and RSMA-enabled NF-ISAC\cite{10906379}. The recent advances in these categories are surveyed next.

\subsubsection{SDMA/NOMA-enabled FF-ISAC} Many studies mainly consider FF-ISAC designs, where the sensing performance is evaluated by  Cram\'{e}r-Rao bound (CRB) \cite{10251151,10382465,10328645}, sensing rate\cite{10571110,10464353,10679658}, or transmit beampattern\cite{10159012}. Specifically,  the authors in \cite{10251151,10382465} study the complete response matrix and reflection angle estimations corresponding to target detection and tracking stages. Imperfect channel state information (CSI) and on-off control of non-transmission power are respectively considered in \cite{10679658} and \cite{10328645} to enhance robustness and energy efficiency. Besides, ISAC can integrate with other technologies. For example, multiple base stations (BS) can be cooperatively scheduled in cell-free networks to provide multi-angle observations and higher spatial diversity \cite{10571110}. These contributions exploit spatial multiplexing gains to counter interference. However, when many users are scheduled per time slot, excess interference saturates the performance of FF-ISAC, where unlimited increases in transmit power fail to enhance transmit rate\cite{9835151}. To elevate performance, the authors in \cite{9668964,10542219} design NOMA-enabled FF-ISACs. However, the limited gains require a stringent decoding order, complex receiver designs, and large channel gain differences.

\subsubsection{RSMA-enabled FF-ISAC} RSMA can overcome rate saturation and reduce receiver complexity. It achieves a higher fairness rate\cite{9991090} and higher sum rate\cite{SumRate} over SDMA and NOMA. Given these benefits, RSMA has been exploited for FF-ISAC networks in \cite{9531484,10486996,10522473,10032141,10287099,10614103}. For example, RSMA is investigated to detect multiple moving targets for mono-static ISAC systems, where a general CRB sensing metric is derived \cite{10486996}. To enhance beampattern matching capacity, reference \cite{9531484} injects an additional radar sequence and utilizes RSMA to control the interference level between the radar sequence and communication streams. Apart from injecting radar sequences, the common stream of RSMA can be independently designed to enhance sensing performance \cite{10522473}. In addition, RSMA-enabled ISAC can be extended to reconfigurable intelligence surface (RIS)\cite{10287099}, cloud radio access networks (C-RAN)\cite{10032141},  and satellite systems\cite{10614103}. However, these efforts, while validating RSMA's efficacy in ISAC, have been confined to the far-field regime. 

\subsubsection{SDMA-enabled NF-ISAC} Currently, NF-ISAC remains largely unexplored, except for \cite{10419546,10520715,10135096,10579914,gavras2024simultaneous}. Communication and dedicated sensing beamforming in single-target scenarios are jointly designed to optimize the beampattern\cite{10419546}. A multi-target detection approach is developed in \cite{10520715}, where the sensing rate is used as the evaluation metric. However, these two works utilize fully digital antennas, requiring high RF costs. To attack this issue, the authors in \cite{10135096} adopt a hybrid array architecture and derive the CRB for the NF joint distance and angle estimation. Furthermore, the authors in \cite{10579914} propose a double-array structure for downlink and uplink ISAC. However, these NF-ISAC works mainly employ SDMA techniques, utilizing beam focusing of the spherical-wave to counter interference\cite{10419546,10520715,10135096,10579914,gavras2024simultaneous}.

\subsubsection{RSMA-enabled NF-ISAC} Although RSMA has been applied in NF communications\cite{10414053,10798456}, its potential in NF-ISAC is underexplored, with only\cite{10906379} beginning to address this gap. In\cite{10906379}, the authors show that spatial beams initially configured for communication users can be repurposed to sense an additional NF target. However, this work does not leverage RSMA to jointly optimize communication and sensing performance. As a result, the question of RSMA’s effectiveness in enhancing network performance is unanswered, an issue our work aims to explore. Furthermore, neither study resolves the key question of whether dedicated sensing beams are necessary. 

\subsection{Contributions}
To tackle the above challenges and the gaps in the literature, this work proposes a novel NF-ISAC transmit scheme with a hybrid beamforming architecture, leveraging RSMA for flexible interference management. We determine whether dedicated sensing beams are necessary for NF multi-target detection, leading to the development of optimized hybrid beamforming algorithms. Table \ref{Table I}  compares our work and existing ISAC studies. Our contributions  are  outlined below:
\begin{table*}[h]
	\caption{Our contributions in contrast to the existing ISAC designs}
    \vspace{-0.3cm}
	\begin{center}\label{Table I}
		\begin{tabular}{|c||c|c|c|c|c|c|} 
			\hline
&\cite{10251151,10382465,10328645,10571110,10464353,10679658,10159012,9668964,10542219} &\cite{9531484,10486996,10522473,10032141,10287099,10614103}&\cite{10419546,10520715}&\cite{10135096,10579914,gavras2024simultaneous}&\cite{10906379}&\makecell*[c]{\bf{Our work}}\\
                \hline 
           	\makecell*[c]{\bf{NF effect}} & \XSolidBrush & \XSolidBrush&\Checkmark&\Checkmark &\Checkmark&\Checkmark\\
                \hline 
           	\makecell*[c]{\bf{RSMA}} & \XSolidBrush & \Checkmark&\XSolidBrush &\XSolidBrush &\Checkmark&\Checkmark\\
                \hline 
           	\makecell*[c]{\bf{Hybrid beamforming }} & \XSolidBrush& \XSolidBrush&\XSolidBrush &\Checkmark &\XSolidBrush&\Checkmark\\
	          \hline 
           	\makecell*[c]{\bf{Uncertain sensing beam count}} & \XSolidBrush&\XSolidBrush & \XSolidBrush&\XSolidBrush &\XSolidBrush  &\Checkmark\\
	          \hline
		\end{tabular}
	\end{center}
\end{table*}
\begin{itemize}
\item A novel RSMA-enabled NF-ISAC system is proposed, where the BS simultaneously serves multiple communication users and detects numerous targets. To achieve this, the system employs a hybrid beamforming architecture. The primary objective is to maximize the minimum communication rate while satisfying transmit power constraints and ensuring the required sensing performance for multi-target detection. This involves the joint optimization of several critical components, including receive filters, the analog beamformer, digital communication and sensing beamformers, common rate allocation, and the allocation of dedicated sensing beams.
\item 
An optimal solution reconstruction approach is developed to ascertain the impact of the number of dedicated sensing beams on the sensing rate. Specifically, an equivalent solution that produces the same objective value can always be constructed with the known optimal digital communication and sensing beamformers. Furthermore, the reconstructed sensing beamformer exhibits a rank-zero structure, indicating that dedicated sensing beams are unnecessary for NF multi-target detection.

\item Building on the above insight, the focus shifts to optimizing analog and digital communication beamformers, receive filters, and common rate allocation. To achieve this, auxiliary variables are introduced, and a penalty dual decomposition (PDD)-based double-loop algorithm is developed. Specifically, the weighted minimum mean-squared error (WMMSE) and quadratic transform methods are employed to recast communication and sensing rates into easily optimized constraints. Then, the introduced auxiliary variable is solved via a convex optimization framework. Additionally, the optimal analog beamformer, digital beamformer, and receive filters are derived with closed-form expressions. 

\item 
Extensive simulations highlight three key advantages of the proposed scheme over five competing benchmarks:

\begin{enumerate}
    \item \textbf{Efficiency in hardware utilization}:  performance comparable to a fully digital beamformer is achieved while significantly reducing the required RF chains.

    \item \textbf{Balanced dual functionality}:  
effective multi-target detection in  NF-ISAC systems is demonstrated without degrading communication performance.

    \item \textbf{Superior performance gains}:  
    The scheme outperforms both SDMA, NOMA, and FF-ISAC approaches, demonstrating substantial improvements in overall system efficiency and capability.
\end{enumerate}
\end{itemize}

\emph{Organization:}  The remainder of this paper is organized as follows.  Section \ref{Section II} elaborates on the system model and formulates the optimization problem. Section \ref{Section III} rigorously proves that dedicated sensing beams are not required for NF multi-target detection. Section \ref{Section IV} presents the proposed iterative optimization algorithm and analyzes its properties.  Section \ref{Section IV} provides simulation results. Section \ref{Section V} concludes this paper. 

\emph{Notations:} Boldface upper-case letters, boldface lower-case letters, and calligraphy letters denote matrices, vectors, and sets, respectively. The $N\times K$ dimensional complex matrix space is denoted by $\mathbb{C}^{N\times K}$. Superscripts ${(\bullet)}^T$ and ${(\bullet)}^H$ represent the transpose and Hermitian transpose, respectively. $\mathrm{Re}\left( \bullet\right)$, $\mathrm{Tr}\left( \bullet\right)$,  $\mathrm{rank}\left( \bullet\right)$, and $\mathbb{E}\left[\bullet\right]$ 
 denote the real part, trace, rank, and statistical expectation, respectively. Operator $\lfloor a\rfloor$  is the largest integer not greater than $a$. $\mathcal{CN}(\mu, \sigma^2)$ denotes a complex Gaussian of mean $\mu$ and variance $\sigma^2$.

\section{System model and problem formulation}\label{Section II}
\begin{figure*}[tbp]
\centering
\includegraphics[scale=0.8]{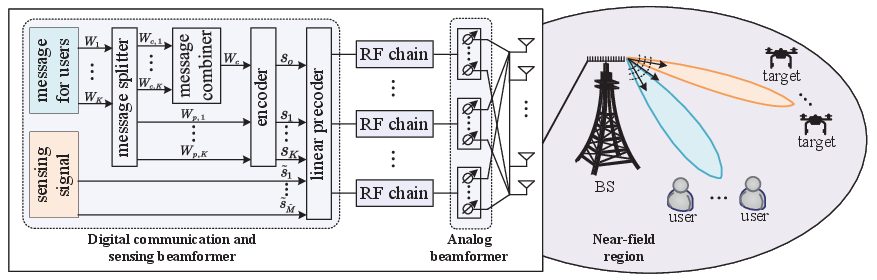}
\caption{The considered RSMA-enabled NF-ISAC networks.}
\label{fig:system}
\end{figure*}
As per Fig.~\ref{fig:system}, an RSMA-enabled NF-ISAC network comprises a dual-functional BS, $K$ single-antenna communication users, and $M$ sensing targets. The sets of communication users and sensing targets are indexed by $\mathcal{K}=\{1,\dots,K\}$ and $\mathcal{M}=\{1,\dots,M\}$, respectively. The BS is equipped with a uniform linear array (ULA) of $N_t$-transmit and $N_r$-receive antennas with an antenna spacing of $d$. The boundary between NF and FF regions is determined by Rayleigh distance $d_i=\frac{2D^2_i}{\lambda}$ for $\forall i\in\{t,r\}$, where $D_i=(N_i-1)d$ and $\lambda$ are antenna aperture and signal wavelength, respectively. All users and targets are assumed to be located in the NF region. In NF-ISAC, fully digital beamforming is impractical since allocating a dedicated RF chain to each antenna is difficult. This paper adopts the hybrid beamforming architecture to attack this issue, as shown in the left half of Fig.~\ref{fig:system}. Specifically, a phase-shifted analog beamformer is placed between  $N_{f}$ ($N_f<N_t$) RF chains and $N_t$ transmit antennas, where the output of each RF chain is sent to all the transmit antennas.

\subsection{Communication and sensing channel models}
Let the reference point of the ULA be located at $\left(0,0\right)$ and the coordinate of the $n$-th transmit antenna is $\mathbf{s}_n=\left(0, nd\right)$, where $n\in\mathcal{N}_t=\{1,\dots,N_t\}$. Consider a user located at $r$ distance and $\theta$ angle from the center of the transmit ULA, so its coordinate is $\mathbf{r}=\left(r\cos\theta,r\sin\theta\right)$. Then, the distance from the $n$-th transmit antenna to this user can be calculated as  
\begin{equation}
d_{n}\left(r,\theta\right)=||\mathbf{r}-\mathbf{s}_n|| = \sqrt{r^2+(nd)^2-2rnd\sin\theta}.
\end{equation}
The resultant corresponding channel, $\tilde h_{n}$, can be modeled as $\tilde h_{n}= \tilde\beta e^{-j\frac{2\pi}{\lambda}d_{n}\left(r,\theta\right)}$, where $\tilde\beta$ and $e^{-j\frac{2\pi}{\lambda}d_{n}\left(r,\theta\right)}$ represent the free-space large-scale path loss and an effect akin to small-scale fading, respectively. In particular, $\tilde \beta=\frac{c}{4\pi fr}$, where $f$ and $c$ are the carrier frequency and speed of light, respectively. To capture the spherical wave characteristic in the NF, the second-order Taylor expansion can be utilized to approximate $d_{n}\left(r,\theta\right)$, \emph{i.e.}, $d_{n}\left(r,\theta\right)\approx r-\delta_{n}\left(r,\theta\right)$, where $\delta_{n}\left(r,\theta\right)=nd\sin\theta-(nd)^2\cos^2\theta/2r$\cite{10517348}. Plugging it into $\tilde h_{n}$, the NF channel between the $n$-th transmit antenna and the user can be rewritten as $h_{n}= \beta e^{j\frac{2\pi}{\lambda}\delta_{n}\left(r,\theta\right)}$, where $\beta=\tilde\beta e^{-j\frac{2\pi}{\lambda}r}$.  Then, the overall NF channel $\mathbf{h}\in\mathbb{C}^{N_t\times 1}$ between the transmit antennas and the user is
\begin{equation}
\mathbf{h}= \beta\left[e^{j\frac{2\pi}{\lambda}\delta_{1}\left(r,\theta\right)},\dots,e^{j\frac{2\pi}{\lambda}\delta_{N_t}\left(r,\theta\right)}\right]^T=\beta\mathbf{a}\left(r,\theta\right).
\label{Channel}
\end{equation}
where $\mathbf{a}\left(r,\theta\right)$ denotes the NF array response vector.

The general multi-path channel model comprises one line-of-sight (LoS) path and $Q$ non-LoS (NLoS) paths induced by $Q$ scatters. Let $r_k$ ($r_{k,q}$) and $\theta_k$ ($\theta_{k,q}$) represent the distance and angle of user $k $ ($q$-th scatter associated to user $k$). For this model, channel $\mathbf{h}_k\in\mathbb{C}^{N_t\times 1}$ can be characterized as
\begin{equation}
\mathbf{h}_k= \beta_k\mathbf{a}\left(r_k,\theta_k\right) + \sum_{q=1}^{Q}\beta_{k,q}\mathbf{a}\left(r_{k,q},\theta_{k,q}\right).
\label{Overall_Channel}
\end{equation}
where $\beta_k=\tilde\beta_k e^{-j\frac{2\pi}{\lambda}r_k}$ and $\beta_{k,q}=\tilde\beta_{k,q} e^{-j\frac{2\pi}{\lambda}\left(r_{k,q}+\tilde r_{k,q}\right)}$ are the complex channel gains of the LoS and the $q$-th NLoS components. $\tilde r_{k,q}$ is the distance between the $k$-th user and the $q$-th scatter. 

Target detecting depends on the echo signal received by the BS, so the BS needs to send probing signals and then gather echo signals. Using the array response vector in equation (\ref{Channel}), round-trip sensing channel matrix $\mathbf{G}_m\in\mathbb{C}^{N_r\times N_t}$ of the $m$-th target can be modeled as \cite{10579914}
\begin{equation}
\mathbf{G}_m=\overline\beta_m\mathbf{b}\left(\overline r_m,\overline\theta_m\right)\mathbf{a}^T\left(\overline r_m,\overline\theta_m\right),
\end{equation}
where $\overline\beta_m$ and $\left(\overline r_m,\overline\theta_m\right)$ are the round-trip complex channel gain and coordinate of the $m$-th target, respectively. $\mathbf{b}\left(\overline r_m,\overline\theta_m\right)\in\mathbb{C}^{N_r\times 1}$ and $\mathbf{a}\left(\overline r_m,\overline\theta_m\right)\in\mathbb{C}^{N_t\times 1}$ denote the receive and transmit NF array response vector, respectively.

\subsection{Signal model, communication rate, and sensing rate}
The BS utilizes downlink RSMA to serve communication users. Specifically, at the encoding level, message $W_k$ for the $k$-th user is split into a common part $W_{c,k}$ and a private part $W_{p,k}$ for $\forall k\in\mathcal{K}$. All common parts $\left\{W_{c,1},\dots,W_{c,K}\right\}$ are combined and encoded into one common stream $s_{0}$ using a public codebook shared by all users. The private parts $\left\{W_{p,1},\dots,W_{p,K}\right\}$ are independently encoded into user-specific streams $\left\{s_{1},\dots,s_{K}\right\}$\footnote{The common part helps converting intra-user interference into a decodable and utilizable signal. In contrast, the private part delivers user-specific information exclusively assigned to each user. By strategically integrating these two components, RSMA provides a more flexible and adaptive interference management framework.}. The unit-power signal stream vector at time index $l$ can be expressed as $\mathbf{s}\left(l\right)=\left[s_0(l),s_1(l),\dots,s_K(l)\right]^T$, where $\forall l\in\mathcal{L}=\{1,\dots, L\}$ is the discrete-time index and $L$ is the total transmit blocks within one coherent processing interval (CPI). The stream vector $\mathbf{s}\left(l\right)$ is linearly precoded by the hybrid beamformer $\mathbf{F}\mathbf{W}\in\mathbb{C}^{N_t\times (K+1)}$ to form communication beams, where $\mathbf{F}\in\mathbb{C}^{N_t\times N_f}$  is the analog beamforming matrix and $\mathbf{W}=\left[\mathbf{w}_0,\mathbf{w}_1,\dots,\mathbf{w}_k\right]\in\mathbb{C}^{N_f\times (K+1)}$ is the digital communication beamforming matrix. $\mathbf{w}_0\in\mathbb{C}^{N_f\times 1}$ and $\mathbf{w}_k\in\mathbb{C}^{N_f\times 1}$ are the beamformers for the common stream and the $k$-th private stream, respectively. Additionally, to enhance sensing performance, the BS injects dedicated sensing beams. Similar to communication beams, these sensing streams are precoded by $\mathbf{F}\mathbf{V}\in\mathbb{C}^{N_t\times N_s}$ for $0\leq N_s\leq N_f$  and then are superimposed with communications beams, where $\mathbf{V}=\left[\mathbf{v}_1,\dots,\mathbf{v}_{N_s}\right]\in\mathbb{C}^{N_f\times N_s}$. Note that, to have the maximum flexibility for ISAC beamforming design,  a variable number of dedicated sensing beams, denoted by $N_s$, is considered. Therefore, the transmitted signal at time index $l$ can be written as 
\begin{equation}
\mathbf{x}\left(l\right)=\mathbf{F}\mathbf{w}_0s_0\left(l\right) + \sum_{k=1}^{K}{\mathbf{F}\mathbf{w}_ks_k\left(l\right)} + \sum_{m=1}^{N_s}{\mathbf{F}\mathbf{v}_m\tilde{s}_m\left(l\right)},
\label{transmit_signal}
\end{equation}
where $\tilde{s}_m\left(l\right)$ is the $m$-th dedicated sensing signal with unit-power at time index $l$. As such, the received signal at the $k$-th user can be written as 
\begin{equation}
y_k\left(l\right)=\mathbf{h}^H_k\mathbf{x}\left(l\right)+n_k,
\end{equation}
where $n_k\sim \mathcal{CN}\left(0,\sigma^2_k\right)$ denotes additional white Gaussian noise (AWGN) term. The average received power for the $k$-th user can be calculated by equation (\ref{equ:rece_power}) as shown at the top of the next page.
\begin{table*}[th]
\hrule
\begin{equation}
T_{c,k}=\overbrace{{\left|\mathbf{h}^H_{k}\mathbf{F}\mathbf{w}_0\right|}^2}^{S_{c,k}}+\underbrace{\overbrace{{\left|\mathbf{h}^H_{k}\mathbf{F}\mathbf{w}_{k}\right|}^2}^{S_{{p,k}}}+\overbrace{\sum_{j=1,j\neq k}^{K}\left|\mathbf{h}^H_{k}\mathbf{F}\mathbf{w}_j\right|^2+\sum_{m=1}^{N_s}\left|\mathbf{h}^H_{k}\mathbf{F}\mathbf{v}_m\right|^2+\sigma^2}^{I_{{p,k}}}}_{I_{c,k}=T_{p,k}}.
\label{equ:rece_power}
\end{equation}
\hrule
\end{table*}

To recover $W_{k}$, the $k$-th user first decodes $s_0$ by treating all private and dedicated sensing streams as noise\cite{mao2018rate}. The common stream $s_0$  includes user ID tags, allowing each user to identify and extract its designated bit segment with a complexity of $\mathcal{O}\left(1\right)$\cite{9831440}. The signal-to-interference-plus-noise ratio (SINR) is $\gamma_{c,k}=S_{c,k}{I^{-1}_{c,k}}$. To ensure that all users can successfully decode $s_0$, the common rate shall not exceed $R_c= \min_{\forall k}R_{c,k}$, where $R_{c,k}=\log\left(1+\gamma_{c,k}\right)$. Additionally, since all users share the common rate,  $R_c=\sum_{k=1}^{K}C_{c,k}$, where $C_{c,k}$ is the portion of the common rate transmitting $W_{c,k}$\cite{9831440}. 
Then, the $k$-th user removes the common stream via successive interference cancellation (SIC) and decodes the desired private stream by treating the residual streams as noise.  The resultant SINR and achievable rate are $\gamma_{{p,k}}=S_{p,k}{I^{-1}_{{p,k}}}$ and $R_{p,k}=\log\left(1 + \gamma_{{p,k}}\right)$, respectively. As such, the total transmit rate of user $k$ is $R_k=C_{c,k}+R_{p,k}$.

Concurrently, the detecting targets reflect $\mathbf{x}\left(l\right)$ to the BS, so the received echo signal at time index $l$ is 
\begin{equation}
\mathbf{y}\left(l\right) =\sum_{m=1}^{M}\sqrt{\alpha_m}\mathbf{G}_m\mathbf{x}\left(l\right)+\mathbf{G}_{SI}\mathbf{x}\left(l\right)+\mathbf{n}_0,
\end{equation}
where $\alpha_m$ is the power reflection coefficient of the $m$-th target, $\mathbf{G}_{SI}\in\mathbb{C}^{N_r\times N_t}$ is self-interference channel due to simultaneous transmission and reception, and $\mathbf{n}_0\sim \mathcal{CN}\left(0,\sigma^2_0\mathbf{I}_{N_r}\right)$ is AWGN. This paper considers complete decoupling between transmit and receive arrays, ensuring that the dual-functional BS remains free from self-interference thanks to self-interference mitigation methods\cite{10679658}. A similar assumption has been made in \cite{10579914,10135096}. After self-interference cancellation, the BS utilizes the receive filter $\mathbf{u}_m\in\mathbb{C}^{N_r\times 1}$ to acquire the desired reflected signal of the $m$-th target. The post-processed signal is thus given as 
\begin{equation}
y(l)=\mathbf{u}^H_m\sum_{m=1}^{M}\sqrt{\alpha_m}\mathbf{G}_m\mathbf{x}\left(l\right)+\mathbf{u}^H_m\mathbf{n}_0.
\end{equation}
The resultant detecting SINR of the $m$-th target is given as
\begin{equation}
\gamma_m=\frac{\alpha_m\mathbf{u}^H_m\mathbf{G}_m\mathbf{R}\mathbf{G}^H_m\mathbf{u}_m}{\mathbf{u}^H_m\left(\sum_{j=1,j\neq m}^{M}\alpha_j\mathbf{G}_j\mathbf{R}\mathbf{G}^H_j+\sigma^2_0\mathbf{I}_{N_r}\right)\mathbf{u}_m},
\label{sensing_rate}
\end{equation}
where $\mathbf{R}=\mathbb{E}\left[\mathbf{x}(l)\mathbf{x}^H(l)\right] =\mathbf{F}\mathbf{W}\mathbf{W}^H\mathbf{F}^H+\mathbf{F}\mathbf{V}\mathbf{V}^H\mathbf{F}^H$ is the covariance matrix. Consequently, the sensing rate of the $m$-th target is $\tilde R_{m}=\log\left(1+\gamma_m\right)$.

\subsection{Problem formulation}
The transmit rate $R_k$ and sensing rate $\tilde R_{m}$ indicate that dedicated sensing beams incur harmful interference to users, while the communication beams can be repurposed to support target detection. In this context, a fundamental question arises: \emph{i.e.}, {\emph{does target detection need dedicated sensing beams? In other words, is NF target detection possible only using communication beams?}} To address this issue, this paper aims to maximize the minimum communication rate among all users while meeting multi-target sensing rate constraints. This process requires the joint optimization of the number of dedicated sensing beams, analog beamformer, digital communication and sensing beamformers, receive filters, and common rate allocation. Therefore, this problem is formulated as
\begin{subequations}\label{linear_p}
	\begin{align}
&\max_{N_s,\mathbf{F},\mathbf{W},\mathbf{U},\mathbf{V},\mathbf{c} } \min_{\forall k} R_k,\label{ob_a}\\
	\text{s.t.}~
	&||\mathbf{F}\mathbf{W}||^2+||\mathbf{F}\mathbf{V}||^2\leq P_{\text{th}},\label{ob_b}\\&
 \tilde R_{m}\geq R_{\text{th}}, \quad \forall m,\label{ob_c}\\
 &\sum_{k=1}^{K}C_{c,k} \leq R_c,\label{ob_d}\\
 &C_{c,k} \geq 0, \quad \forall k,\label{ob_e}\\
  &|\mathbf{F}_{i,j}|=1,~1\leq i\leq N_t,~1\leq j\leq N_f,\label{ob_f}\\  
  &||\mathbf{u}_m||^2=1,\quad\forall m,\label{ob_g}\\
  &1\leq N_s\leq N_f,\label{ob_h}
	\end{align}
\end{subequations}
where $\mathbf{U}=[\mathbf{u}_1,\dots,\mathbf{u}_M]$ and $\mathbf{c}=[C_{c,1},\dots, C_{c,K}]$. $P_{\text{th}}$ and $R_{\text{th}}$ denote the maximum transmit power and the sensing rate requirement thresholds, respectively.  (\ref{ob_b}) and (\ref{ob_c}) are the transmit power and sensing performance constraints, respectively.  (\ref{ob_d}) and (\ref{ob_e}) are the common rate allocation constraints. (\ref{ob_f}) is the unit-modulus constraint of the analog beamformer while (\ref{ob_g}) is the receive filter normalization constraint. (\ref{ob_h}) is the number constraint of dedicated sensing beams. 

Problem (\ref{linear_p}) appears elusive to optimally solve due to three technical challenges. First, the logarithmic function in $R_k$ and $\tilde R_m$ and minimum operator in $R_{c}$ incur non-convexity and non-smoothness. Such problems are difficult to solve in both primal and dual domains since the duality gap is unknown. Second, the analog beamformer, digital beamformers, and receive filters are intricately coupled and cannot be separated, aggravating the solution difficulty. Third, the number of dedicated sensing beams is uncertain, complicating the optimization process. Consequently, the optimal solution is intractable.

\section{Dedicated sensing beams or not?}\label{Section III}
This section aims to ascertain the number of dedicated sensing beams, eliminating its adverse impact in solving the problem (\ref{linear_p}). For this, the intuitive idea is to get the optimal solution and then check whether equation  $N_s=0$ holds. However, as discussed earlier, the optimal solution to problem (\ref{linear_p}) is mathematically intractable. An optimal solution reconstruction method is proposed to address this challenge, which rigorously meets $N_s=0$. This reveals that dedicated sensing beams are not required for NF multi-target detection.

To eliminate the uncertainty incurred by $N_s$, we introduce auxiliary matrices $\tilde{\mathbf{W}}_k=\mathbf{w}_k\mathbf{w}^H_k$, $\tilde{\mathbf{ V}}=\sum_{m=1}^{N_s}\mathbf{v}_m\mathbf{v}^H_m$, $\tilde{\mathbf{h}}_k=\mathbf{F}^H\mathbf{h}_k$, and $\mathbf{H}_k=\tilde{\mathbf{h}}_k\tilde{\mathbf{h}}^H_k$. As a result, the number of dedicated sensing beams is determined by the rank of $\tilde{\mathbf{ V}}$, \emph{i.e.}, $N_s = \mathrm{rank}\big(\tilde{\mathbf{V}}\big)$. Then, problem (\ref{linear_p}) can be recast as
\begin{subequations}\label{linear_p3}
	\begin{align}
&\max_{N_s,\mathbf{F},\tilde{\mathbf{W}}_k,\tilde{\mathbf{V}}, \mathbf{U}, \mathbf{c} } \min_{\forall k} R_k,\label{ob_a3}\\
	\text{s.t.}~&\sum_{k=0}^{K}\mathrm{Tr}\left(\mathbf{F}^H\mathbf{F}\tilde{\mathbf{W}}_k\right)+\mathrm{Tr}\left(\mathbf{F}^H\mathbf{F}\tilde{\mathbf{V}}\right)\leq P_{\text{th}},\label{ob_b3}\\
 &\tilde{\mathbf{W}}_k\succeq \mathbf{0},~\tilde{\mathbf{V}}\succeq \mathbf{0},~\forall k\in\mathcal{K}_1,\label{ob_c3}\\
&\mathrm{rank}\big(\tilde{\mathbf{W}}_k\big)=1,~\forall k\in\mathcal{K}_1,\label{ob_d3}\\
 &\mbox{ (\ref{ob_c}) -- (\ref{ob_h})}.\label{ob_e3}
	\end{align}
\end{subequations}
where $\mathcal{K}_1=\left\{0,1,\dots,K\right\}$. Note that equation (\ref{equ:rece_power}) should be substituted by equation (\ref{equ:rece_power2}) shown at the top of the next page in calculating the communication rate. \begin{table*}[th]
\hrule
\begin{align}
T_{c,k}=\overbrace{\mathrm{Tr}\left(\mathbf{H}_k\tilde{\mathbf{W}}_0\right)}^{S_{c,k}}+
\underbrace{\overbrace{\mathrm{Tr}\left(\mathbf{H}_k\tilde{\mathbf{W}}_k\right)}^{S_{{p,k}}}+\overbrace{\sum_{j=1,j\neq k}^{K}\mathrm{Tr}\left(\mathbf{H}_k\tilde{\mathbf{W}}_j\right)+\mathrm{Tr}\left(\mathbf{H}_k\tilde{\mathbf{ V}}\right)+\sigma^2}^{I_{{p,k}}}}_{I_{c,k}=T_{p,k}}.
\label{equ:rece_power2}
\end{align}
\hrule
\end{table*}
Meanwhile, $\mathbf{R}$ should be updated to $\mathbf{R}=\sum_{k=0}^{K}\mathbf{F}\tilde{\mathbf{W}}_k\mathbf{F}^H+\mathbf{F}\tilde{\mathbf{V}}\mathbf{F}^H$. The non-convex rank-one constraint and fractional SINR make the optimization problem intractable for direct solutions. Therefore,  the specific value of the optimal solution cannot be determined, but there should be at least one optimal solution.  Let $\mathcal{Q}^*_1 = \big\{\mathbf{F}^*,\tilde{\mathbf{W}}^*_k,\tilde{\mathbf{V}}^*, \mathbf{U}^*, \mathbf{c}^*\big\}$ denote the optimal solution, but $\mathrm{rank}\big({\tilde{\mathbf{V}}^*}\big)$ is difficult to be determined analytically. To counter this challenge,  Proposition 1, proved in Appendix A, constructs an equivalent optimal rank-zero solution but may violate $\mathrm{rank}\big(\tilde{\mathbf{W}}_k\big)=1$.

{\bf\emph{Proposition 1}}:  If rank-one constraint (\ref{ob_d3}) is temporarily relaxed, one can always construct other solution $ \mathcal{Q}_2= \big\{\mathbf{F}^*,\hat{\mathbf{W}}_k,\hat{\mathbf{V}}, \mathbf{U}^*, \mathbf{c}^*\big\}$  that guarantees $E\left(\mathcal{Q}_2\right)\geq E\left(\mathcal{Q}^*_1\right)$, where $E\left(\mathcal{Q}\right)$  denotes the objective value achieved by the solution $\mathcal{Q}$ and 
\begin{equation}
\hat{\mathbf{W}}_k=\tilde{\mathbf{W}}^*_k+\delta_k \tilde{\mathbf{V}}^* ~\mathrm{and} ~ \hat{\mathbf{V}}=\mathbf{0}.
\label{Equaivalent}
\end{equation}
In equation (\ref{Equaivalent}), the arbitrary weighted coefficients must meet $\sum_{k=0}^{K}\delta_k=1$ and $\delta_k\geq 0$.

Proposition 1 reveals that removing dedicated sensing beams will not hinder the sensing rate when neglecting the rank-one constraints (\ref{ob_d3}). However, the constructed digital communication beamformers may not be rank-one, \emph{i.e.}, $ \hat{\mathbf{W}}_k\geq 1$, so the conclusion in Proposition 1 cannot apply to the problem (\ref{linear_p3}). Proposition 2, proved in Appendix B, creates a feasible rank-one solution from the high-rank solution $\hat{\mathbf{W}}_k$ to address this issue.

{\bf\emph{Proposition 2}}: Keeping other variable blocks unchanged, a feasible rank-one solution $\big\{\bar{\mathbf{W}}_k=\bar{\mathbf{w}}_k\bar{\mathbf{w}}^H_k\big\}$  can be constructed for problem (\ref{linear_p3}), which yields the same performance with high-rank solution $\big\{\hat{\mathbf{W}}_k\big\}$, \emph{i.e.}, $E\left(\mathcal{Q}_3\right)=E\left(\mathcal{Q}_2\right)$, where $\mathcal{Q}_3=\big\{\mathbf{F}^*,\bar{\mathbf{W}}_k,\hat{\mathbf{V}}, \mathbf{U}^*, \mathbf{c}^*\big\}$. 

Proposition 2 proves that $\mathcal{Q}_3$ can satisfy all of the constraints in (\ref{linear_p3}). Combining Proposition 1 and Proposition 2 yields 
\begin{equation}
E\left(\mathcal{Q}_3\right)=E\left(\mathcal{Q}_2\right)\geq E\left(\mathcal{Q}^*_1\right).
\label{Performance}
\end{equation}
Equation (\ref{Performance}) shows that $\mathcal{Q}_3$ can yield the same performance with  $\mathcal{Q}^*_1$ at least.  Since $\mathcal{Q}^*_1$ has produced the maximal objective value, it is thus deduced that the equation sign holds and $\mathcal{Q}_3$ is an optimal solution.  Therefore, there exists an optimal solution making $\mathbf{V}=0$, which indicates the non-necessity of dedicated sensing beams in NF multi-target detection, \emph{i.e.}, $N_s^*=0$.

\textbf{Remark 1}: Our findings align with previous works in  ISAC\cite{9531484,liu2020joint,chen2020composite}. Specifically, the authors in\cite{chen2020composite} reveal that a dedicated sensing beam is superfluous for single-user scenarios. Similarly, reference\cite{9531484} illustrates that the RSMA-enabled transmit scheme with and without dedicated sensing beams achieves the same tradeoff performance. Nonetheless, this assertion is substantiated solely by simulation results, lacking the support of rigorous theoretical analysis.  These studies provide indirect corroboration for the validity of our theoretical proof.  Besides, references\cite{liu2020joint,chen2020composite} indicate that employing dedicated probing signals can enhance sensing performance for multi-user ISAC. However, performance gains are contingent upon certain conditions, including the mitigation of interference from dedicated sensing beams at the user end or the implementation of zero-forcing beamforming techniques. Therefore, our conclusion is not in conflict with these two works.

\section{Algorithm and analysis}\label{Section IV}
This section focuses on optimizing the analog beamformer, digital communication beamformers, receive filters, and common rate allocation, given that dedicated sensing beams are unnecessary. To optimize the system, the PDD, WMMSE\cite{8846761}, and quadratic transform (as first proposed in \cite{shen2018fractional}) approaches are leveraged to design a PDD-based double-loop algorithm. Specifically, the PDD method is used to reformulate the problem (\ref{linear_p}) into a more manageable form. The WMMSE and quadratic transform methods are then applied to recast communication and sensing rates into easily optimized constraints. Additionally, the algorithm is summarized, and its convergence and complexity are discussed.

To proceed, we introduce an auxiliary matrix $\mathbf{P}=\left[\mathbf{p}_0,\dots,\mathbf{p}_k\right]\in\mathbb{C}^{N_t\times (K+1)}$, which meets $\mathbf{p}_k=\mathbf{F}\mathbf{w}_k$ for $\forall k\in\mathcal{K}_1$. Plugging $N_s^*=0$ and $\mathbf{p}_k=\mathbf{F}\mathbf{w}_k$ into equation (\ref{equ:rece_power}), the received power by the $k$-th user can be adjusted to 
\begin{equation}
T_{c,k}=\overbrace{{\left|\mathbf{h}^H_{k}\mathbf{p}_0\right|}^2}^{S_{c,k}}+\underbrace{\overbrace{{\left|\mathbf{h}^H_{k}\mathbf{p}_{k}\right|}^2}^{S_{{p,k}}}+\overbrace{\sum_{j=1,j\neq k}^{K}\left|\mathbf{h}^H_{k}\mathbf{p}_j\right|^2+\sigma^2}^{I_{{p,k}}}}_{I_{c,k}=T_{p,k}},
\label{equ:rece_power3}
\end{equation}
which is used to recalculate the common and private rates. Similarly, the covariance matrix should be updated to $\mathbf{R} =\mathbf{P}\mathbf{P}^H$. We introduce a non-negative auxiliary variable $R_s$ to attack the non-smoothness incurred by (\ref{ob_a}). Consequently, the resultant new problem is
\begin{subequations}\label{linear_p4}
	\begin{align}
&\max_{\mathcal{Q}} R_s,\label{ob_a4}\\
	\text{s.t.}~
	&\mathbf{P}=\mathbf{F}\mathbf{W},\label{ob_b4}\\
 &||\mathbf{P}||^2\leq P_{\text{th}},\label{ob_c4}\\
 &R_k\geq R_s,~\forall k,\label{ob_d4}\\
 &\mbox{ (\ref{ob_c}) -- (\ref{ob_g})}.\label{ob_f4}
	\end{align}
\end{subequations}
where $\mathcal{Q}=\left\{\mathbf{P},\mathbf{F},\mathbf{W},\mathbf{U}, \mathbf{c}, R_s \right\}$ collects all optimization variables. The equality constraint (\ref{ob_a4}) still hinders the optimization process. To address this issue, the PDD approach\cite{9120361} is adopted.  Specifically, it transfers constraint (\ref{ob_b4}) into the objective function by introducing the Lagrangian dual variable and penalty parameter, creating a double-loop iterative problem. The outer loop updates the Lagrangian dual matrix and the penalty parameter. Readers are referred to \cite{9120361} for more details about updating rules. The inner loop solves the augmented Lagrangian (AL) problem. By introducing Lagrangian dual matrix $\mathbf{D}$ and penalty parameter $\rho$, the AL problem is formulated as
\begin{subequations}\label{linear_p5}
	\begin{align}
&\max_{\mathcal{Q}} R_s - \mathbf{D}^H\left(\mathbf{P}-\mathbf{FW}\right)-\frac{1}{2\rho}||\mathbf{P}-\mathbf{FW}||^2,\label{ob_a5}\\
	\text{s.t.}~
 &\mbox{ (\ref{ob_c}) -- (\ref{ob_g}),  (\ref{ob_c4}), (\ref{ob_d4})}.\label{ob_b5}
	\end{align}
\end{subequations}
Alg.~\ref{Alg.1} summarizes the proposed PDD-based double-loop algorithm framework.  Given the known dual matrix and penalty parameter, solving problem (\ref{linear_p5})   appears intractable. However, the constraints in problem (\ref{linear_p5}) are separable. This observation motivates us to divide variables into several blocks and optimize each block alternately. The optimization of each block is elaborated next.
\begin{algorithm}[t]
	\caption{PDD-based framework for solving problem (\ref{linear_p4})}
	\begin{algorithmic}[1]\label{Alg.1}
		\STATE Initialize $\mathbf{F}^{(0)}$, $\mathbf{W}^{(0)}$, $\mathbf{D}^{(0)}$, $\rho^{(0)}$, and $\psi^{(0)}$, set iteration index $n=1$ and the maximum tolerance $\xi_1$.
		\WHILE{ not convergent}
		\STATE  Solving problem (\ref{linear_p5}) to obtain $\mathbf{P}^{(n)}$, $\mathbf{F}^{(n)}$, and $\mathbf{W}^{(n)}$  via Alg.~\ref{Alg.2}.
        \IF{ $||\mathbf{P}^{(n)}-\mathbf{F}^{(n)}\mathbf{W}^{(n)}||_\infty\leq \psi^{(n-1)}$ }
		\STATE Update $\mathbf{D}^{(n)}=\mathbf{D}^{(n-1)}+\frac{1}{\rho^{(n)}}\left(\mathbf{P}^{(n)}-\mathbf{F}^{(n)}\mathbf{W}^{(n)}\right)$.
        \STATE Keep penalty factor unchanged, \emph{i.e.},~$\rho^{(n)}=\rho^{(n-1)}$.
        \ELSE
        \STATE Update $\rho^{(n)}=\mu\rho^{(n-1)}$, where $0<\mu<1$.
        \STATE Keep Lagrangian dual matrix $\mathbf{D}$ unchanged, \emph{i.e.}, $\mathbf{D}^{(n)}=\mathbf{D}^{(n-1)}$.
		\ENDIF
        \STATE Update $\psi^{(n)}=0.9||\mathbf{P}^{(n)}-\mathbf{F}^{(n)}\mathbf{W}^{(n)}||_\infty$ and $n=n+1$.
		\ENDWHILE	
		\STATE Output the maximized minimum sensing rate.
	\end{algorithmic}
\end{algorithm}

\subsection{Subproblem with respect to $\{\mathbf{U}\}$}
With fixed auxiliary matrix $\mathbf{P}$, optimizing $\mathbf{u}_m$ for detecting the $m$-th target will not impact other targets. Therefore, problem (\ref{linear_p5}) can be divided into $M$ independent subproblems. In each subproblem, $\mathbf{u}_m$ is optimized to enhance the sensing SINR for the $m$-th target, which yields  the following optimization problem
\begin{subequations}\label{linear_p6}
	\begin{align}
&\max_{\mathbf{u}_m}\frac{\alpha_m\mathbf{u}^H_m\mathbf{G}_m\mathbf{R}\mathbf{G}^H_m\mathbf{u}_m}{\mathbf{u}^H_m\mathbf{Q}_m\mathbf{u}_m}\label{ob_a6}\\
	\text{s.t.}~
	&||\mathbf{u}_m||^2=1,\label{ob_b6}
	\end{align}
\end{subequations}
where $\mathbf{Q}_m=\sum_{j=1,j\neq m}^{M}\alpha_j\mathbf{G}_j\mathbf{R}\mathbf{G}^H_j+\sigma^2_0\mathbf{I}_{N_r}$. Problem (\ref{linear_p6}) is a generalized Rayleigh quotient problem\cite{10571110}, and its optimal solution is 
\begin{equation}
\mathbf{u}^*_m = \mathbf{v}_{max}\left(\mathbf{Q}^{-1}_m\mathbf{G}_m\mathbf{R}\mathbf{G}^H_m\right), ~\forall m,
\label{Optimal_filters}
\end{equation}
where $\mathbf{v}_{max}(\mathbf{A})$ denotes the eigenvector corresponding to the maximum eigenvalue of matrix $\mathbf{A}$.

\subsection{Subproblem with respect to $\big\{\mathbf{P},\mathbf{c},R_s\big\}$}
With fixed analog beamformer $\mathbf{F}$, digital beamformer $\mathbf{W}$, and receive filter $\mathbf{U}$, problem (\ref{linear_p5})  can be simplified as
\begin{subequations}\label{linear_p7}
	\begin{align}
&\max_{\mathbf{P},\mathbf{c},R_s} R_s - \mathbf{D}^H\left(\mathbf{P}-\mathbf{FW}\right)-\frac{1}{2\rho}||\mathbf{P}-\mathbf{FW}||^2,\label{ob_a7}\\
	\text{s.t.}~
 &\mbox{ (\ref{ob_c}), (\ref{ob_d}), (\ref{ob_e}), (\ref{ob_c4}), (\ref{ob_d4})}.\label{ob_b7}
	\end{align}
\end{subequations}

\emph{1) WMMSE for communication rate:} Problem (\ref{linear_p7}) involves the fractional SINR and coupled auxiliary variables, which is difficult to solve directly. The WMMSE approach is particularly effective in dealing with logarithmic transmit rate expressions by introducing equalizers\cite{8846761}. Thus,  it is employed to recast common rate $R_{c,k}$ and private rate $R_{p,k}$. Specifically, the $k$-th user utilizes equalizer $\omega_{c,k}$ to the received signal, realizing the estimation of $s_{0}$, denoted by $\hat{s}_{c,k}=\omega_{c,k}y_k$, where time index $l$ is dropped for brevity. After removing the common stream via SIC, equalizer $\omega_{p,k}$ is applied to obtain an estimate of $\hat{s}_{k}$ given by $\hat{s}_{k}=\omega_{p,k}\left(y_k-\mathbf{h}^H_k\mathbf{p}_0s_0\right)$. Subsequently, the mean-squared errors (MSEs) of common and private streams, defined respectively as $\delta_{c,k}=\mathbb{E}\left\{\left|\hat{s}_{c,k}-s_{0}\right|^2\right\}$ and $\delta_{p,k}=\mathbb{E}\left\{\left|\hat{s}_{k}-s_{k}\right|^2\right\}$, are given as
\begin{subequations}\label{MMSE_error}
\begin{align}
\delta_{c,k} = &\left|\omega_{c,k}\right|^2T_{c,k}-2\mathrm{Re}\left(\omega_{c,k}\mathbf{h}^H_k\mathbf{p}_0\right)+1,\\
\delta_{p,k} = &\left|\omega_{p,k}\right|^2T_{p,k}-2\mathrm{Re}\left(\omega_{p,k}\mathbf{h}^H_k\mathbf{p}_k\right)+1.
\end{align}
\end{subequations}
By solving  $\frac{\partial \delta_{c,k}}{\partial \omega_{c,k}}=0$ and $\frac{\partial \delta_{p,k}}{\partial \omega_{p,k}}=0$, the optimum minimum MSE (MMSE) equalizers are respectively expressed as 
\begin{align}
\omega^{\mathrm{MMSE}}_{c,k}=\mathbf{p}^H_0\mathbf{h}_kT^{-1}_{c,k} \quad \mathrm{and }\quad\omega^{\mathrm{MMSE}}_{p,k}=\mathbf{p}^H_k\mathbf{h}_kT^{-1}_{p,k}.
\label{Optimal_equalizer}
\end{align}
Substituting (\ref{Optimal_equalizer}) into (\ref{MMSE_error}), the resulting MMSEs are written as
\begin{subequations}
\begin{align}
\delta^{\mathrm{MMSE}}_{c,k} =\min_{\omega_{c,k}}\delta_{c,k}= &T^{-1}_{c,k}I_{c,k},\\
\delta^{\mathrm{MMSE}}_{p,k} =\min_{\omega_{p,k}}\delta_{p,k}= &T^{-1}_{p,k}I_{p,k}.
\label{MMSE_error2}
\end{align}
\end{subequations}
The MMSEs are related to the SINRs such that $\gamma_{c,k}=1/\delta^{\mathrm{MMSE}}_{c,k}-1$ and $\gamma_{p,k}=1/\delta^{\mathrm{MMSE}}_{p,k}-1$, from which the transmit rate can be expressed as $R_{c,k}=-\log\big(\delta^{\mathrm{MMSE}}_{c,k}\big)$ and $R_{p,k}=-\log\big(\delta^{\mathrm{MMSE}}_{p,k}\big)$, respectively.

Next,  the augmented weighted MSEs (WMSEs) are defined as $\beta_{c,k}=\eta_{c,k}\delta_{c,k}-\log\left(\eta_{c,k}\right)$ and $\beta_{p,k}=\eta_{p,k}\delta_{p,k}-\log\left(\eta_{p,k}\right)$, where $\eta_{c,k}$ and $\eta_{p,k}$ are the weights associated with the $k$-th user's MSEs. By taking
the equalizers and weights as optimization variables, the rate--WMMSE relationship is established as 
\begin{subequations}\label{Relationship}
\begin{align}
\beta^{\mathrm{MMSE}}_{c,k} &=\min_{\eta_{c,k},\omega_{c,k}}\beta_{c,k}= \tau-R_{c,k},\\
\beta^{\mathrm{MMSE}}_{p,k} &=\min_{\eta_{p,k},\omega_{p,k}}\beta_{p,k}= \tau-R_{p,k},
\end{align}
\end{subequations}
where $\tau=1/\ln 2 +\log(\ln 2)$. By setting $\frac{\partial \beta_{c,k}}{\partial \omega_{c,k}}=0$ and $\frac{\partial \beta_{p,k}}{\partial \omega_{p,k}}=0$, the optimal equalizers become $\omega^*_{c,k}=\omega^{\mathrm{MMSE}}_{c,k}$ and $\omega^*_{p,k}=\omega^{\mathrm{MMSE}}_{p,k}$. Similarly, the optimal weights can be derived as follows 
\begin{align}
\eta^*_{c,k}=\left(\delta^{\mathrm{MMSE}}_{c,k}\ln 2\right)^{-1}~\mathrm{and} ~\eta^*_{p,k}=\left(\delta^{\mathrm{MMSE}}_{p,k}\ln 2\right)^{-1}.
\label{Optimal_weights}
\end{align}
By closely examining each WMSE, it is convex in each variable when the other two are specified.

\emph{2) Quadratic transform for sensing rate:} 
The sensing rate's numerator depends on the covariance matrix $\mathbf{R}$, a function of all beamforming vectors. Due to this coupling, it is challenging to derive a suitable equalizer that could reformulate the sensing rate into a tractable form. Consequently, the WMMSE approach cannot transform the sensing rate, necessitating an alternative solution. Here, the quadratic transform approach is employed to construct an accurate surrogate for the sensing rate. Proposition 3 is motivated by Theorem 2 in \cite{shen2018fractional}.

{\bf\emph{Proposition 3}}: After introducing our constructed surrogate function $f\left(\mathbf{x},\mathbf{P}\right) = 2\mathrm{Re}\left(\mathbf{x}^{H}\mathbf{s}\left(\mathbf{P}\right)-\mathbf{x}^{H}I\left(\mathbf{P}\right)\mathbf{x}\right)$ for any $\mathbf{s}(\mathbf{P})\in\mathbb C^{N_t\times 1}$ and $I(\mathbf{P}) > 0$, we can derive 
\begin{equation}
\frac{(\mathbf{s}\left(\mathbf{P}\right))^H\mathbf{s}\left(\mathbf{P}\right)}{I\left(\mathbf{p}\right)}=\max_{\mathbf{x}}f\left(\mathbf{x},\mathbf{P}\right) 
\label{Surrogate}
\end{equation} 
and the optimal solution to the right-hand of equation (\ref{Surrogate}) is $\mathbf{x}^*=\frac{\mathbf{s}\left(\mathbf{P}\right)}{I\left(\mathbf{P}\right)}$.

Engaging Proposition 3 to recast the sensing SINR yields 
\begin{align}
f_{m}\left(\mathbf{x}_{m},\mathbf{P}\right)= 2\alpha_m\mathrm{Re}\left(\mathbf{x}^H_{m}\mathbf{u}^H_{m}\mathbf{q}_{m}\right)-\mathbf{x}^H_{m}\mathbf{u}^H_m\mathbf{Q}_m\mathbf{u}_m\mathbf{x}_{m},\label{surrogate}
\end{align}
where $\mathbf{x}_m$ is the introduced auxiliary variable. Using the rate-WMMSE relationship (\ref{Relationship}) and constructed surrogate (\ref{surrogate}) for sensing SINR, one can recast (\ref{linear_p7}) as
\begin{subequations}\label{linear_p8}
	\begin{align}
&\max_{\hat{\mathcal {Q}}_1,\hat{\mathcal {Q}}_2} R_s - \mathbf{D}^H\left(\mathbf{P}-\mathbf{FW}\right)-\frac{1}{2\rho}||\mathbf{P}-\mathbf{FW}||^2,\label{ob_a8}\\
	\text{s.t.}~
 &\sum_{k=1}^{K}C_{c,k} + \min_{\eta_{c,k},\omega_{c,k}}\beta_{c,k} \leq \tau,~\forall k,\label{ob_b8}\\
 &C_{c,k} - \min_{\eta_{p,k},\omega_{p,k}}\beta_{p,k} \geq R_{s}-\tau,~\forall k,\label{ob_c8}\\
&\log\left(1+\max_{\mathbf{x}_m}f_m\left(\mathbf{x}_{m},\mathbf{P}\right)\right)\geq R_{\text{th}},~\forall m,\label{ob_d8}\\
 &\mbox{ (\ref{ob_e}), (\ref{ob_c4})}.\label{ob_e8}
	\end{align}
\end{subequations}
$\hat{\mathcal {Q}}_1$ and $\hat{\mathcal {Q}}_2$ collect respectively the intrinsic variables and introduced auxiliary variables, \emph{i.e.}, $\hat{\mathcal {Q}}_1=\big\{\mathbf{P},\mathbf{c},R_s\big\}$ and $\hat{\mathcal {Q}}_2=\big\{\eta_{c,k}, \omega_{c,k}, \eta_{p,k}, \omega_{p,k}, \mathbf{x}_m\big\}$. The non-trivial coupling between $\hat{\mathcal {Q}}_1$ and $\hat{\mathcal {Q}}_2$ makes problem (\ref{linear_p8}) difficult to solve directly. However,  observe that problem (\ref{linear_p8}) becomes convex upon fixing introduced auxiliary variable $\hat{\mathcal {Q}}_2$. Driven by this observation, we consider optimizing $\hat{\mathcal {Q}}_i$ while keeping $\hat{\mathcal {Q}}_j$ at its previous value and vice versa, where $\forall i,j\in\{1,2\}$ and $j\neq i$. Specifically, the intrinsic variables are solved via convex optimization solvers while the optimal auxiliary variables with closed-form expressions are derived according to  (\ref{Optimal_equalizer}), (\ref{Optimal_weights}), and  (\ref{close-form}).  That is,  
\begin{equation}\label{close-form}
\mathbf{x}^*_m =\mathbf{u}^H_{m}\mathbf{q}_{m},~\forall m, 
\end{equation}
which is derived from Proposition 3.

\subsection{Subproblem with respect to $\big\{\mathbf{F}, \mathbf{W}\big\}$}
The variables $\mathbf{F}$ and $\mathbf{W}$ only appear in the last two-term of the objective function, so problem (\ref{linear_p5}) reduces to
\begin{subequations}\label{linear_p9}
	\begin{align}
&\min_{\mathbf{F},\mathbf{W} } \mathbf{D}^H\left(\mathbf{P}-\mathbf{FW}\right)+\frac{1}{2\rho}||\mathbf{P}-\mathbf{FW}||^2,\label{ob_a9}\\
	\text{s.t.}~
 &\mbox{ (\ref{ob_f})}.\label{ob_b9}
	\end{align}
\end{subequations}
To streamline the optimization process, (\ref{ob_a9}) is equivalently transformed to $\frac{1}{2\rho}||\mathbf{P}-\mathbf{FW}+\rho\mathbf{D}||^2 + \rho^2\mathbf{D}^2$. Omitting constant $\rho^2\mathbf{D}^2$ and positive coefficient $\frac{1}{2\rho}$, problem (\ref{linear_p9}) is recast to 
\begin{subequations}\label{linear_10}
	\begin{align}
&\min_{\mathbf{F},\mathbf{W} } ||\mathbf{P}-\mathbf{FW}+\rho\mathbf{D}||^2,\label{ob_a10}\\
	\text{s.t.}~
 &\mbox{ (\ref{ob_f})}.\label{ob_b10}
	\end{align}
\end{subequations}
 As (\ref{linear_p9}) is a highly coupled quadratic problem, optimizing analog and digital beamformers alternately is the solution. 

\emph{1) Digital beamformer optimization:} \blue{} With fixed $\mathbf{F}$, problem (\ref{linear_p9}) is simplified as $\min_{\mathbf{W}}||\mathbf{P}-\mathbf{FW}+\rho\mathbf{D}||^2$, which is a quadratic function. By solving $\frac{\partial ||\mathbf{P}-\mathbf{FW}+\rho\mathbf{D}||^2}{\partial \mathbf{W}}=0$, the optimal solution is found as   
\begin{equation}
\mathbf{W}^*=\left(\mathbf{F}^H\mathbf{F}\right)^{-1}\mathbf{F}^H\left(\mathbf{P}+\rho\mathbf{D}\right).
\label{Optimal_digital}
\end{equation}

\emph{2) Analog beamformer optimization:} With fixed $\mathbf{W}$, problem (\ref{linear_p9}) can be simplified as
\begin{subequations}\label{linear_p10}
	\begin{align}
&\min_{\mathbf{F} }\mathrm{Tr}\left(\mathbf{F}^H\mathbf{F}\mathbf{Y}\right)-2\mathrm{Re}\left(\mathrm{Tr}\left(\mathbf{F}^H\mathbf{Z}\right)\right),\label{ob_a10}\\
	\text{s.t.}~
 &|\mathbf{F}_{i,j}|=1,~1\leq i\leq N_t,~1\leq j\leq N_f,\label{ob_b10}
	\end{align}
\end{subequations}
where $\mathbf{Y}=\mathbf{W}\mathbf{W}^H$ and $\mathbf{Z}=\left(\mathbf{P}+\rho\mathbf{D}\right)\mathbf{W}^H$. The elements of $\mathbf{F}$ are separated in unit-modulus constraint (\ref{ob_b10}), which motivates us to optimize $\mathbf{F}$ by one element at a time. Consequently, the optimization problem for  $\mathbf{F}_{i,j}$  reduces to 
\begin{subequations}\label{linear_p11}
	\begin{align}
&\min_{\mathbf{F}_{i,j}}\phi_{i,j}|\mathbf{F}_{i,j}|^2-2\mathrm{Re}\left(\chi_{i,j}\mathbf{F}_{i,j}\right),\label{ob_a11}\\
\text{s.t.}~
 &\mbox{ (\ref{ob_b10})}.\label{ob_b11}
	\end{align}
\end{subequations}
where $\phi_{i,j}$ and $\chi_{i,j}$ are real and complex constant coefficients determined by the elements of $\mathbf{F}$ except for $\mathbf{F}_{i,j}$, respectively. Under unit-modulus constraint (\ref{ob_b10}), optimal $\mathbf{F}_{i,j}$ can be obtained by
\begin{equation}
\mathbf{F}^*_{i,j}=\frac{\chi^H_{i,j}}{|\chi_{i,j}|}.
\label{Optimal_analog}
\end{equation} 
At present, coefficient  $\chi_{i,j}$ in (\ref{linear_p11}) remains unknown. However, objective functions (\ref{linear_p10}) and (\ref{linear_p11}) have the same partial derivatives with respect to $\mathbf{F}_{i,j}$, so we can derive 
\begin{equation}
\mathbf{X}_{i,j} -\mathbf{Z}_{i,j} = \phi_{i,j}\tilde{\mathbf{F}}_{i,j}-\chi_{i,j},
\end{equation} 
where $\mathbf{X} = \tilde{\mathbf{F}}\mathbf{Y}$ and $\tilde{\mathbf{F}}$ denotes that the optimized solution of $\mathbf{F}$ in the previous iteration. Moreover, expanding $\tilde{\mathbf{F}}\mathbf{Y}$ yields $\phi_{i,j}\tilde{\mathbf{F}}_{i,j}=\tilde{\mathbf{F}}_{i,j}\mathbf{Y}_{j,j}$, so we have 
\begin{equation}
\chi_{i,j}=\mathbf{Z}_{i,j}-\mathbf{X}_{i,j} + \tilde{\mathbf{F}}_{i,j}\mathbf{Y}_{j,j}.
\end{equation} 

The proposed alternating algorithm for problem (\ref{linear_p5}) is summarized in Alg.~\ref{Alg.2}, where receive filters, digital beamformer, analog beamformer, and introduced auxiliary variables are iteratively updated till convergence.

\textbf{Remark 2:} Our algorithm exhibits two important extensibility properties: 1) compatibility with three-dimensional (3D) coordinates and 2) robustness to imperfect SIC. In the 3D extension, the distance between the $n$-th transmit antenna and users is determined by both zenith and azimuth angles while maintaining the same mathematical formulation as in the 2D case. For imperfect SIC scenarios, the residual common stream $\alpha S_{c,k}$ introduces additional interference during private stream decoding,  modifying the SINR to $\gamma_{{p,k}}=S_{p,k}{(\alpha S_{c,k} + I_{{p,k}})^{^{-1}}}$, where $\alpha$ is the residual percentage. By redefining the interference term as $\tilde{I}_{p,k}=\alpha S_{c,k} + I_{{p,k}}$ and substituting ${I}_{p,k}$ with $\tilde{I}_{p,k}$ in our proposed algorithm, the framework maintains full functionality without structural modifications.

\begin{algorithm}[t]
	\caption{Alternating algorithm for solving problem (\ref{linear_p5})}
	\begin{algorithmic}[1]\label{Alg.2}
		\STATE Set $\mathbf{F}^{(1)}=\mathbf{F}^{(n-1)}$ and $\mathbf{W}^{(1)}=\mathbf{W}^{(n-1)}$, set iteration index $i=1$,  and the maximum tolerance $\xi=10^{-3}$.
		\WHILE{ not convergent}
		\STATE  Update $\mathbf{U}^{(i)}$ according to equation (\ref{Optimal_filters}).
        \STATE Update $\omega^{(i)}_{c,k}$ and $\omega^{(i)}_{p,k}$ according to equation (\ref{Optimal_equalizer}).
        \STATE Update $\eta^{(i)}_{c,k}$ and $\eta^{(i)}_{p,k}$ according to equation (\ref{Optimal_weights}).
        \STATE Update $\mathbf{x}^{(i)}_m$  according to equation (\ref{close-form}).
        \STATE Obtain optimal $\mathbf{P}^{(i)}$ and $\mathbf{c}^{(i)}$ by solving problem (\ref{linear_p8}).
        \STATE Update $\mathbf{W}^{(i+1)}$ according to equation (\ref{Optimal_digital}).
        \STATE Update $\mathbf{F}^{(i+1)}$ according to equation (\ref{Optimal_analog}).
		\STATE Update iteration index $i=i+1$.
		\ENDWHILE	
		\STATE Output $\mathbf{P}^{(n)}$, $\mathbf{F}^{(n)}$, and $\mathbf{W}^{(n)}$.
	\end{algorithmic}
\end{algorithm}

\subsection{Convergence and Complexity Analysis} 
\begin{figure}[tbp]
	\centering
	\includegraphics[scale=0.35]{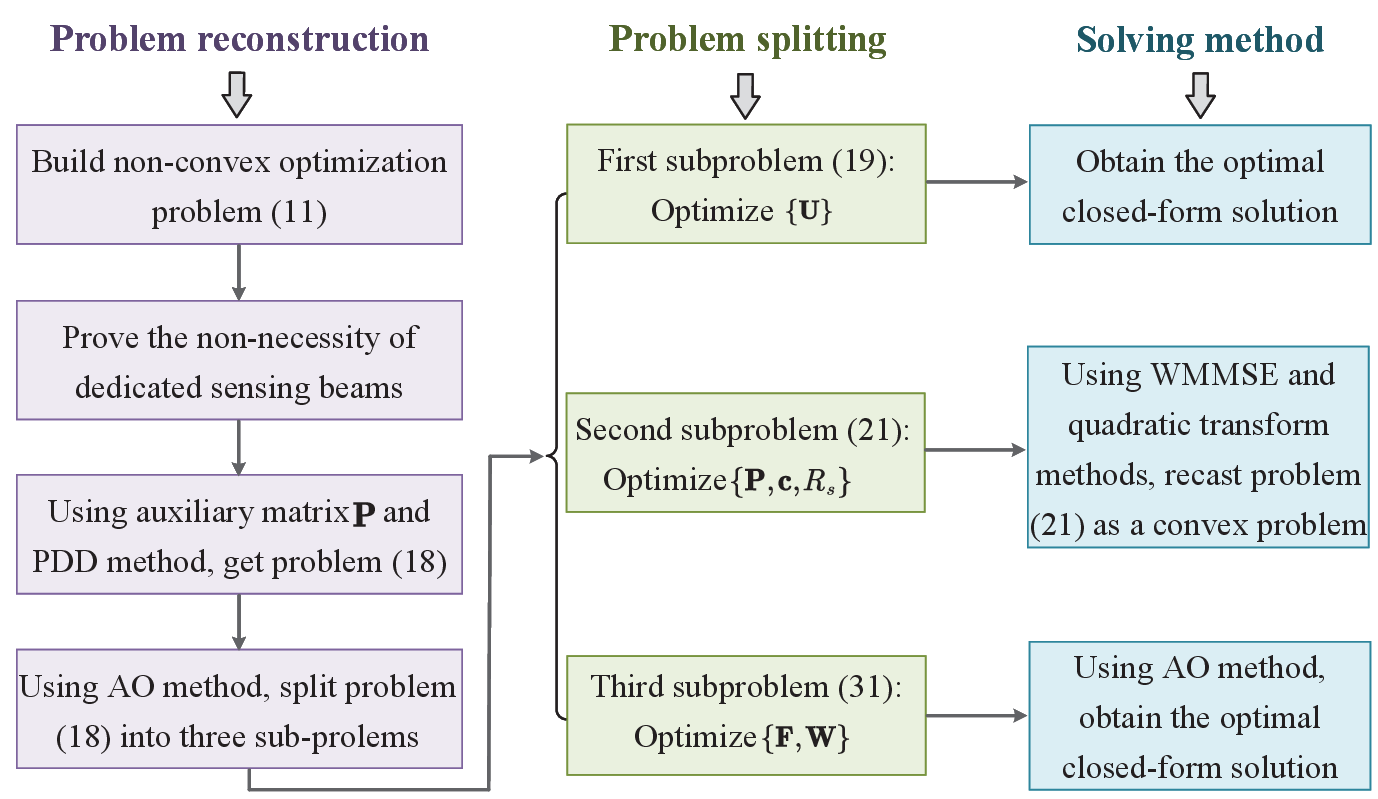}
	\caption{The block diagram of our algorithm, where these three subproblems are iteratively solved until convergence.}
	\label{diagram}
\end{figure}

The proposed solution for (\ref{linear_p4}) is summarized in Alg.~\ref{Alg.1}. Next, its critical properties (\emph{i.e.}, convergence, optimality, and complexity) are discussed.
\begin{itemize}
\item  \emph{Convergence}: Starting from any feasible initial point, the algorithm always yields globally optimal solutions, ensuring it can locate the final feasible point. As a result, the objective value remains stable or improves with each iteration, which means that
\begin{align}\label{convergence}
&R\left(\mathcal{Q}^{(n)}_1,\mathcal{Q}^{(n)}_2,\mathcal{Q}^{(n)}_3\right)\leq R\left(\mathcal{Q}^{(n+1)}_1,\mathcal{Q}^{(n)}_2,\mathcal{Q}^{(n)}_3\right)\notag\\
  &\leq R\left(\mathcal{Q}^{(n+1)}_1,\mathcal{Q}^{(n+1)}_2,\mathcal{Q}^{(n)}_3\right)\\
  &\leq R\left(\mathcal{Q}^{(n+1)}_1,\mathcal{Q}^{(n+1)}_2,\mathcal{Q}^{(n+1)}_3\right)\notag
\end{align}
where $R$ and superscript $n$ denote the objective value and iteration index, respectively. $\mathcal{Q}_1$, $\mathcal{Q}_2$, and $\mathcal{Q}_3$ collect the variables in the three subproblems, respectively. Additionally, since the sensing rate is an upper bound, it is clear that the algorithm converges to a stationary point within a finite number of iterations.

\item \emph{Complexity}: The main computational load in each iteration stems from Alg.~\ref{Alg.2}.  In Alg.~\ref{Alg.2}, optimal $\mathbf{P}$ and $\mathbf{c}$ are solved via CVX while the remaining variables are updated by the closed-form solutions. The complexity using the interior point method is $\mathcal O\left(N_v^{3.5}\right)$, where $N_v$ is the number of variables. Thus, the complexity of updating $\mathbf{P}$ and $\mathbf{c}$ is in order of $\mathcal O\left((K+1)^{3.5}(N_t+1)^{3.5}\right)$. The complexity of updating other variables stems from the matrix inversion, multiplication, and eigenvalue decomposition. For two matrices $\mathbf{W}_1\in\mathbb C^{ A_1\times A_2}$ and $\mathbf{W}_2\in\mathbb C^{ A_2\times A_3}$, the complexity of $\mathbf{W}_1\mathbf{W}_2$ is $\mathcal O\left(A_1A_2A_3\right)$. For matrix $\mathbf{W}_3\in\mathbb C^{ A_4\times A_4}$, the complexity inversion and eigenvalue decomposition are $\mathcal O\left(A^3_4\right)$. Therefore, the complexity from line 3 to line 9 (excluding line 7) are in order of $\mathcal O\left(M\left(N^3_r+N_r^2N_t+N_rN^2_t\right)\right)$, $\mathcal O\left(KN^2_t\right)$, $\mathcal O\left(KN^2_t\right)$, $\mathcal O\left(K\left(N_rN_t+N^2_t\right)\right)$,
$\mathcal O\left(N^2_fN_t+N^3_f+N_fN_T(K+1)\right)$, and $\mathcal O\left(N_tN_f(K+1)\right)$, respectively. By retaining the higher-order terms, the per-iteration computational complexity of Alg.~\ref{Alg.2} is $\mathcal O\left(M\left(N^3_r+N_r^2N_t+N_rN^2_t\right) +(K+1)^{3.5}(N_t+1)^{3.5}\right)$. The algorithm diagram is
given in Fig.~\ref{diagram} to show the procedures of solving the problem (\ref{linear_p}).
\end{itemize}

\section{Simulation result}\label{Section V}
\begin{table}[t]
	\caption{Key simulation parameters}
    \vspace{-0.2cm}
	\begin{center}\label{Table III}
		\begin{tabular}{l l  p{13cm}}       
			\toprule  	\cmidrule(r){1-2}
			$\mathbf{Parameter}$ & $\mathbf{Value}$ \\
			\midrule
			\hline
			Number of transmit antennas $N_t$& $64$\\
			\hline
                Number of receive antennas $N_r$ & $64$\\
			\hline
			Operating frequency $f_c$ & $30$~GHz\\
			\hline
			antenna array aperture $D_t=D_r$ & $0.5$~meter\\
			\hline
			Rayleigh distance $d_t=d_r$ & $50$~meter\\
			\hline
			Number of RF chains $N_f$ & $8$\\
                \hline
                Number of scatters per user $Q$ & $2$\\
                \hline
                Number of communication users $K$ & $6$\\
                \hline
                Number of sensing targets $M$ & $4$ \\
			\hline
			Maximum transmit power $P_{\text{th}}$ & $30$~dBm\\	
			\hline
			Background noise power $\sigma^2$ & $-80$~dBm\\
			\hline
   			Sensing rate requirement $R_{\text{th}}$ & $6$~bps/Hz\\
			\hline
                Power reflection coefficient $\alpha_m$ & 1\\
			\hline
		\end{tabular}
	\end{center}
    \vspace{-0.5cm}
\end{table}

Numerical results evaluate our proposed transmit scheme and algorithm. Scatters for communication links are randomly generated within the distance from $20$~m to $30$~m. Communication users and sensing targets are randomly distributed throughout the NF region.  Unless specified otherwise, Table \ref{Table III} provides all the simulation parameters. These parameters are primarily sourced from \cite{10579914,10520715,10458958}.

Each point is averaged over 100 independent channel realizations. The proposed transmit scheme (labeled as {\bf{RSMA-HB, near}}) is compared against five baselines to assess its performance comprehensively. They are described next. 
\begin{itemize}
\item {\bf{RSMA-FD, near}}: 
Each antenna connects to a dedicated RF chain, enabling full-dimensional beamforming. This fully digital antenna architecture is an upper performance bound \cite{10436390} for the proposed architecture, which adopts an analog and digital beamforming hybrid.  The analog beamformer is constrained by unit modulus, e.g.,  (\ref{ob_f}). In contrast, the full digital beamforming benchmark does not have this constraint, resulting in a larger feasible region.
\item {\bf{NOMA-HB, near}}: Each user's message is encoded to a private stream. The stronger user then successively decodes and cancels the weaker users' streams before decoding its desired stream. 
\item {\bf{SDMA-HB, near}}: The encoding method is the same as the second benchmark, but each user directly decodes its desired stream. Thus,  this baseline disables the common stream, \emph{i.e.}, $\mathbf{w}_0=\mathbf{0}$. 
\item {\bf{RSMA-comm, near}}: This benchmark neglects the requirement for multi-target detection, \emph{i.e.}, $R_{\text{th}}=0$. This reveals the impact of using communication beams to detect targets on communication performance.

\item {\bf{RSMA-HB, far}}: This one adopts the FF channel model to confirm the benefits of NF beamforming. For a fair comparison, all parameters are identical to the NF counterpart except for the array response vector. In FF channels, the array response vector in equation (\ref{Channel}) is updated to
\begin{equation}
\mathbf{a}_{\text{far}}\left(\theta\right)= \left[e^{j\frac{2\pi}{\lambda}d\sin\theta},\dots,e^{j\frac{2\pi}{\lambda}Nd\sin\theta}\right]^T.
\label{Far-Channel}
\end{equation}
\end{itemize}

\begin{figure}[tbp]
	\centering
	\includegraphics[scale=0.45]{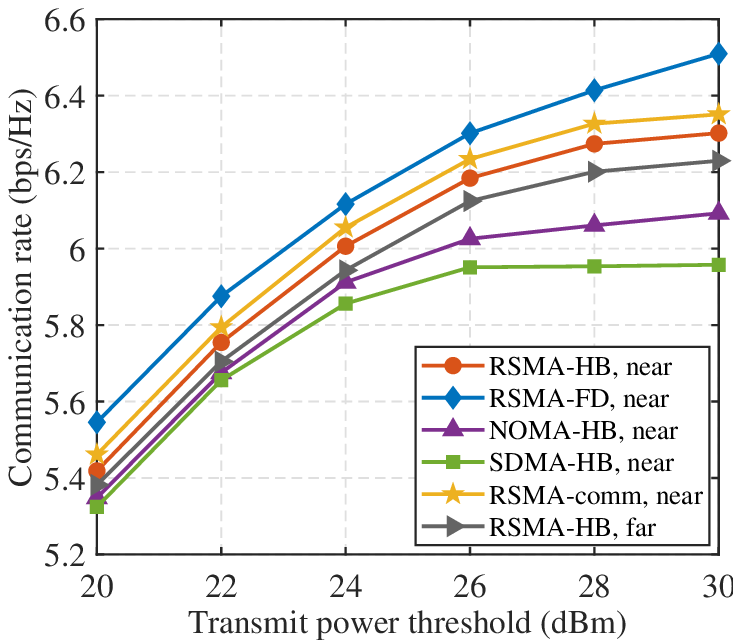}
	\caption{Max-min communication rate versus transmit power.}
	\label{power}
\end{figure}

Figure~\ref{power} shows the max-min communication rate versus transmit power, highlighting four key advantages of the proposed scheme over competing benchmarks:
\begin{enumerate}
    \item \textbf{Effective interference management:} While the communication rate increases with transmit power across all schemes, SDMA stagnates at \(P_{\text{th}} \geq 26\)~dBm due to excessive interference. In contrast, the gap between RSMA and conventional schemes (SDMA and NOMA) widens, demonstrating RSMA’s superior interference handling.
    \item \textbf{NF beamforming superiority:} The proposed NF scheme outperforms the FF system by leveraging NF beamforming to focus energy on specific points and minimize inter-user interference through reduced leakage. 
    \item \textbf{Minimal impact of sensing constraints:} Meeting the sensing rate of \(6\)~bps/Hz reduces the communication rate by only \(0.05\)~bps/Hz, highlighting the efficiency of the RSMA-enabled NF-ISAC design.
    \item \textbf{Comparable to full digital beamforming:} Full digital beamforming achieves a communication rate gain of \(0.05\)~bps/Hz at \(P_{\text{th}} \leq 28\)~dBm and \(0.13\)~bps/Hz at \(P_{\text{th}} = 30\)~dBm over proposed hybrid beamforming. This demonstrates the proposed scheme’s ability to perform near the level of fully digital beamforming. 
\end{enumerate}
\begin{figure}[tbp]
	\centering
	\includegraphics[scale=0.45]{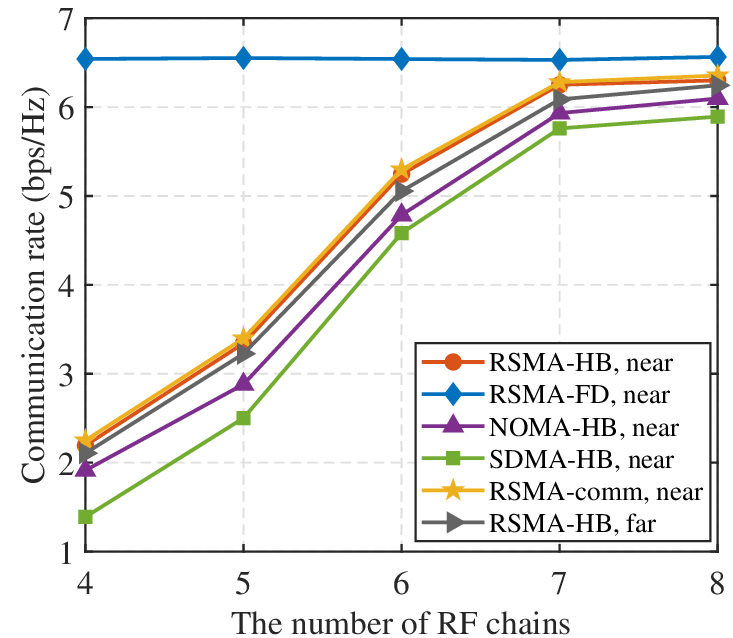}
	\caption{Max-min communication rate versus the number of RF chains.}
	\label{radio}
\end{figure}

Fig.~\ref{radio} shows the max-min communication rate versus the number of RF chains. From the figure, our proposed hybrid beamforming algorithm can achieve a comparable communication rate to the full digital beamforming when $N_f>K$.  This is because that digital beamforming can create sufficient spatial Degrees of Freedom (DoF) to neutralize interference. However, when the number of RF chains decreases, data streams exceed the DoF created through classical digital beamforming, degrading the communication performance. Therefore, the performance gap between hybrid beamforming and full digital beamforming becomes clearer when $N_f<K$. However, under an identical number of RF chains, our proposed transmit scheme always performs close to communication-only networks and surpasses SDMA, NOMA, and FF-ISAC.

\begin{figure}[tbp]
	\centering
	\includegraphics[scale=0.45]{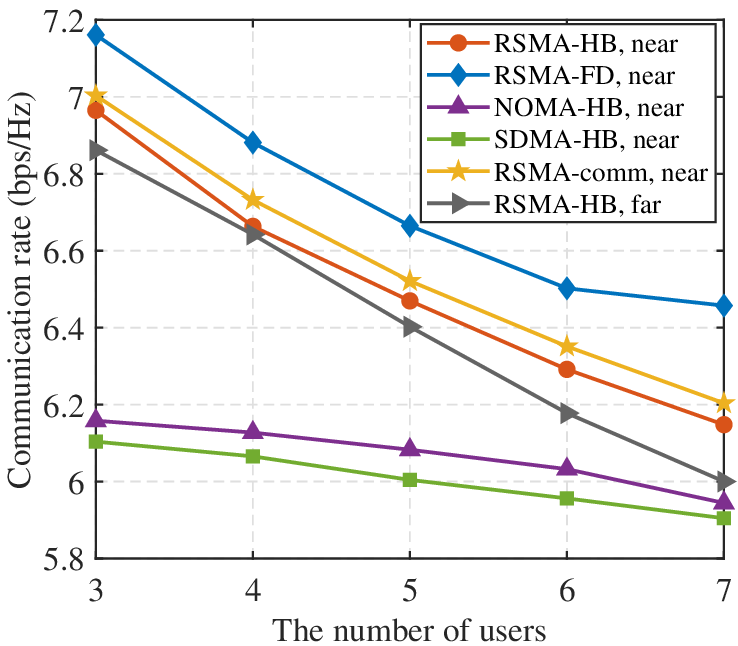}
	\caption{Max-min communication rate versus the number of users.}
	\label{user}
\end{figure}
Fig.~\ref{user} presents the max-min communication rate versus the number of users. As anticipated, all approaches reduce the communication rate as more users are scheduled. Moreover, the performance gap between RSMA and conventional schemes gradually narrows. This phenomenon happens since all users are required to decode $s_0$. However, the common rate depends on the user with the poorest worst channel quality and is shared by all communication users. The gap between NF-ISAC and FF-ISAC becomes more obvious. This is because spherical-wave propagation can distinguish users with similar angular directions and then focus the beam energy on a specific point, which helps neutralize intra-user interference.

\begin{figure}[tbp]
	\centering
	\includegraphics[scale=0.45]{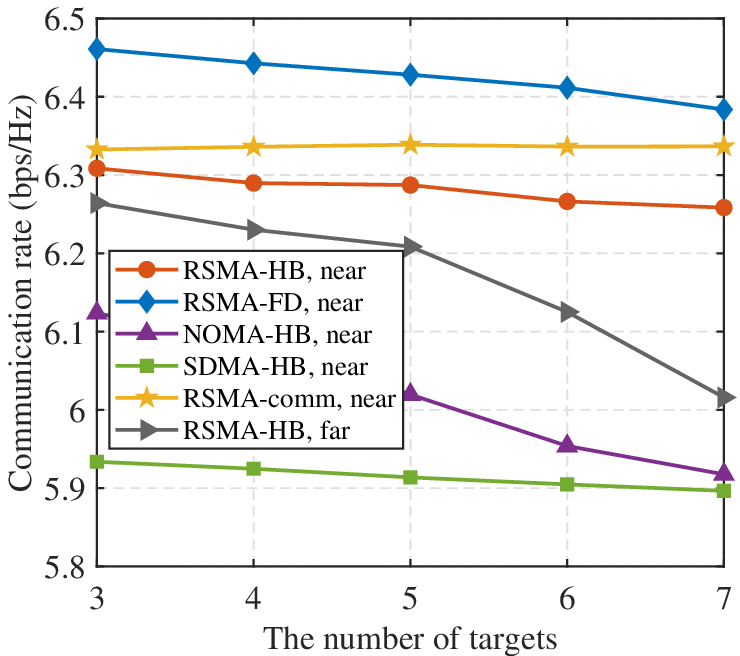}
	\caption{Max-min communication rate versus the number of targets.}
	\label{target}
\end{figure}

Fig.~\ref{target} investigates the impact of the number of targets on the communication rate. All transmit schemes degrade the max-min communication rate as the number of targets increases, except for the communication-only scheme, where the communication rate remains static at approximately $6.33$~bps/Hz. This is because its performance is independent of the target numbers. Interestingly, the FF-ISAC exhibits a significant performance decline beyond five targets, rendering a more pronounced performance gap. For example, the performance gap between NF- and FF-ISACs reaches about $0.25$~bps/Hz when $M=7$. This is because targets with the same angle and distance are more likely to appear as the number of targets increases. Compared to plane-wave propagation, spherical-wave propagation encompasses distance and angle information, which can better cope with such scenarios. In addition, our proposed scheme consistently outperforms SDMA and NOMA, highlighting its effectiveness. 

\begin{figure}[tbp]
	\centering
	\subfigure{
		\begin{minipage}[t]{0.45\linewidth}
			\centering\includegraphics[width = 1.5in]{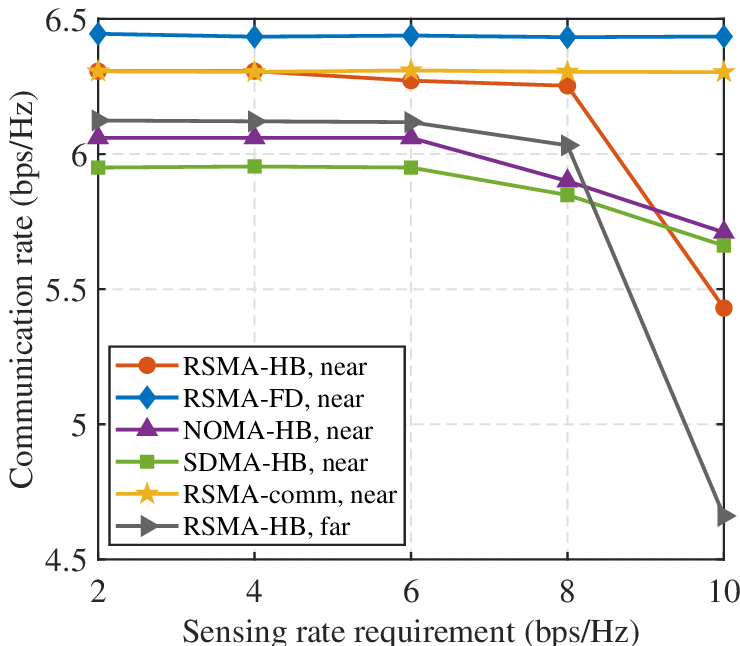}
		\end{minipage}
	}
	\subfigure{
		\begin{minipage}[t]{0.48\linewidth}
			\centering\includegraphics[width = 1.51in]{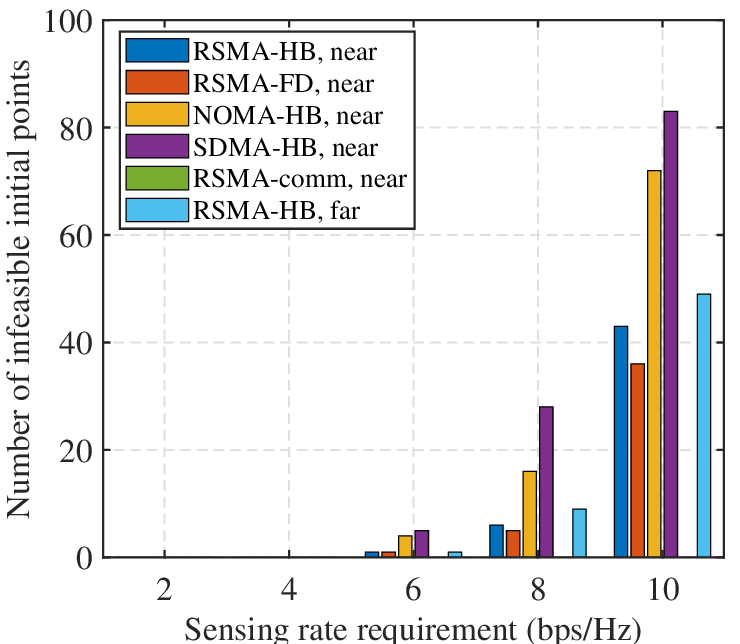}
		\end{minipage}
	}
	\centering
	\caption{(a) Max-min communication rate versus sensing rate. (b) The number of infeasible points versus the sensing rate.}
	\label{QoS}
\end{figure}
Fig.~\ref{QoS}(a) illustrates the max-min communication rate versus sensing rate requirements. At the same time, Fig.~\ref{QoS}(b) shows the number of infeasible points, reflecting instances where the sensing rate requirement is unmet across 100 channel realizations. The communication rate of all transmit schemes decreases slightly or remains static for $R_{\text{th}}<8$~bps/Hz, indicating that multi-target detection requirements are easily supported.  

Two key observations emerge for $R_{\text{th}}=10$~bps/Hz:  
\begin{itemize}
    \item Our scheme outperforms SDMA and NOMA for $R_{\text{th}}<8$~bps/Hz but shows the opposite trend at $R_{\text{th}}=10$~bps/Hz. This occurs because Fig.~\ref{QoS}(b) indicates that SDMA and NOMA fail to meet the sensing rate requirement in 83\% and 72\% of cases at $R_{\text{th}}=10$~bps/Hz, respectively. This lower feasible probability may overstate its performance since infeasible points, often caused by complex channel states, are discarded. 
    \item Our transmit scheme achieves a lower infeasible probability than FF-ISAC, with a performance gain of 0.8~bps/Hz, highlighting the advantages of NF-ISAC.
\end{itemize}

\begin{figure}[tbp]
	\centering
	\includegraphics[scale=0.48]{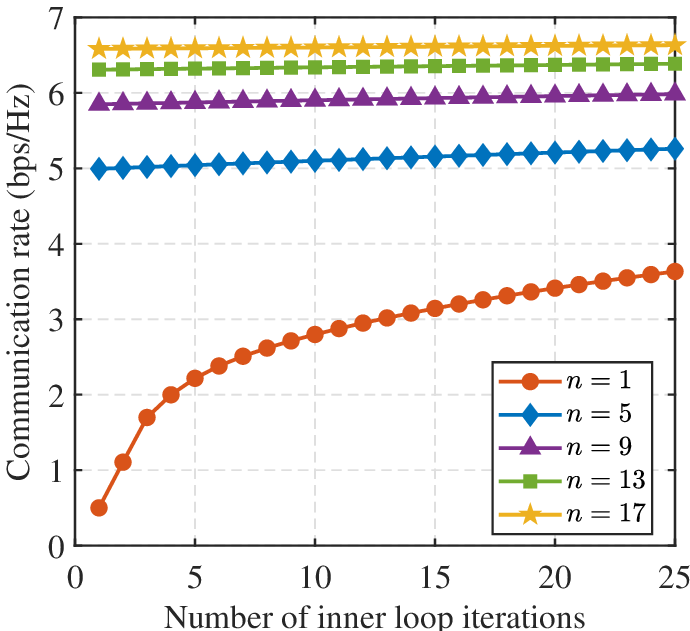}
	\caption{Max-min communication rate versus the number of iterations, where $n$ denotes the number of outer loop iterations.}
	\label{convergence}
\end{figure}
Fig.~\ref{convergence} illustrates the convergence of the proposed PDD-based double-loop algorithm. A steady value is reached after around 17 outer loop iterations, validating the convergence analysis in Section~\ref{Section IV}. The inner loop for solving the AL problem (\ref{linear_p5}) reaches a stationary value within a few iterations when the number of outer loop iterations exceeds five.
\vspace{-0.2cm}
\section{Conclusion}\label{Section VI}
This paper proposes an NF-ISAC RSMA scheme to support multiple users and sensing simultaneously. The receive filters, analog beamformer, digital communication/sensing beamformers, common rate allocation, and the number of dedicated sensing beams are jointly optimized to maximize the minimum communication rate. However, the uncertain sensing beam count and coupled beamformers make the formulated problem discrete and non-convex.
To address the high complexity of this problem, a divide-and-conquer approach is proposed. First, a solution reconstruction method is designed to demonstrate that no dedicated sensing beams are required for NF multi-target detection. Second, a PDD-based double-loop algorithm is developed to optimize the remaining variables jointly. Using WMMSE  and quadratic transform techniques, communication and sensing rates are reformulated, enabling closed-form solutions for the analog beamformer, digital beamformer, and receive filters, while the auxiliary variable is efficiently solved via convex optimization.
Simulation results show that the proposed scheme performs comparably to the fully digital beamformer and communication-only networks while achieving significant gains over other benchmarks.

Our contributions provide valuable insights for future research, highlighting two key directions. First, the challenge of accurate channel estimation in NF-ISAC, which is more complex than in FF-ISAC, presents an opportunity to leverage RSMA’s robustness to mitigate imperfect CSI. Second, NF-ISAC with beam split effect or space non-stationary channels suggests a promising avenue for addressing such scenarios.

\appendices
\section{Proof of Proposition 1}
The latest constructed digital beamformers yield 
\begin{equation}
\sum_{k=0}^{K}\hat{\mathbf{W}}_k = \sum_{k=0}^{K}\tilde{\mathbf{W}}^*_k+\tilde{\mathbf{V}}^*,
\label{Updated_probing}
\end{equation}
which ensures the covariance matrix of the transmit signal remains static, yielding the same sensing rate as $\big\{\tilde{\mathbf{W}}^*_k,\tilde{\mathbf{V}}^*\big\}$.  Additionally, one has
\begin{align}
&\sum_{k=0}^{K}\mathrm{Tr}\left({(\mathbf{F}^*)}^H\mathbf{F}^*\hat{\mathbf{W}}_k\right) = \sum_{k=0}^{K}\mathrm{Tr}\left({(\mathbf{F}^*)}^H\mathbf{F}^*\left(\tilde{\mathbf{W}}^*_k+\delta_k\tilde{\mathbf{V}}^*\right)\right)\notag\\
&\overset{(a)}{=}\sum_{k=0}^{K}\mathrm{Tr}\left(({\mathbf{F}^*)}^H\mathbf{F}^*\tilde{\mathbf{W}}^*_k\right)+\mathrm{Tr}\left({(\mathbf{F}^*)}^H\mathbf{F}^*\tilde{\mathbf{V}}^*\right),
\label{Power_2}
\end{align}
where equation (a) holds since $\sum_{k=0}^{K}\delta_k=1$. As such, the constructed solution meets the transmit power constraint (\ref{ob_b3}) since $\big\{\mathbf{F}^*,\tilde{\mathbf{W}}^*_k,\tilde{\mathbf{V}}^*\big\}$ is a feasible point to problem (\ref{linear_p3}). Then, plugging the constructed solution $\big\{\hat{\mathbf{W}}_k,\hat{\mathbf{V}}\big\}$ into $\gamma_{c,k}$, one can derive
\begin{align}
\hat{\gamma}_{c,k}=&\frac{\mathrm{Tr}\left(\mathbf{H}_k\left(\tilde{\mathbf{W}}^*_0+\delta_0\tilde{\mathbf{V}}^*\right)\right)}{\sum_{j=1}^{K}\mathrm{Tr}\left(\mathbf{H}_k\left(\tilde{\mathbf{W}}^*_j+\delta_j\tilde{\mathbf{V}}^*\right)\right)+\sigma^2} \notag\\\geq& \frac{\mathrm{Tr}\left(\mathbf{H}_k\tilde{\mathbf{W}}^*_0\right)}{\sum_{j=1}^{K}\mathrm{Tr}\left(\mathbf{H}_k\left(\tilde{\mathbf{W}}^*_j+\delta_j\tilde{\mathbf{V}}^*\right)\right)+\sigma^2}\notag\\\geq& \frac{\mathrm{Tr}\left(\mathbf{H}_k\tilde{\mathbf{W}}^*_0\right)}{\sum_{j=1}^{K}\mathrm{Tr}\left(\mathbf{H}_k\left(\tilde{\mathbf{W}}^*_j+\tilde{\mathbf{V}}^*\right)\right)+\sigma^2}=\gamma_{c,k}
\label{Rate}
\end{align}
where the equal sign holds when $\delta_0=0$ and $\hat{\gamma}_{c,k}$ denotes the updated SINR. Similarly, it can be proven $\hat{\gamma}_{p,k}\geq\gamma_{p,k}$, where $\delta_0=0$ and $\delta_k=0$ contributes to the equal sign. It thus can be deduced there is at least one $k$ making $\hat{\gamma}_{p,k}>\gamma_{p,k}$ since $\sum_{k=0}^{K}\delta_k=1$. This indicates that the constructed solution helps elevate the communication rate, so the objective value cannot be decreased and $\mathcal{Q}_2$  can satisfy constraints (\ref{ob_d}) and (\ref{ob_e}).  Moreover, it is easy to validate that the constructed solution meets constraints (\ref{ob_f}), (\ref{ob_g}), (\ref{ob_h}), and (\ref{ob_c3}). Consequently, $\mathcal{Q}_2$ is feasible to the problem (\ref{linear_p3}) with relaxed rank-one constraint. Combining equation (\ref{Updated_probing}), (\ref{Power_2}), and (\ref{Rate}), we deduce that the constructed solution can reach the same objective value as the optimal solution at least.

\section{Proof of Proposition 2}
Without loss generality, assuming $\mathrm{rank}\big(\hat{\mathbf{W}}_k\big)=A_k>1$, one can derive $\hat{\mathbf{W}}_k=\sum_{i=1}^{A_k}\hat{\mathbf{w}}_{k,i}\hat{\mathbf{w}}_{k,i}=\hat{\mathbf{P}}_k\hat{\mathbf{P}}^H_k$ with $\hat{\mathbf{P}}_k=\big[\hat{\mathbf{w}}_{k,1},\dots,\hat{\mathbf{w}}_{k,A_k}\big]\in\mathbb{C}^{N_f\times A_k}$. Define  a Hermitian matrix $\mathbf{X}_k\in\mathbb{C}^{A_k\times A_k}$, which lies in the left null space of $\mathbf{B}_k=\big[\hat{\mathbf{P}}^H_k\hat{\mathbf{B}}_1\hat{\mathbf{P}}_k,\dots, \hat{\mathbf{P}}^H_k\hat{\mathbf{B}}_{M^2+K}\hat{\mathbf{P}}_k\big]\in\mathbb{C}^{A_k\times (M^2+K)A_k}$, \emph{i.e.}, $\mathrm{Tr}\big(\mathbf{B}_k\mathbf{X}_k\big)=0$, where 
\begin{equation}\label{fra}
\hat{\mathbf{B}}_{i}=
\begin{cases}
\mathbf{G}^H_a\mathbf{u}_b\mathbf{u}^H_b\mathbf{G}_a,&\mbox{if $i\leq M^2$ };\\
\mathbf{H}_{i-M^2}, &\mbox{if $i>M^2$ },
\end{cases}
\end{equation}
where $a=\lfloor i/M\rfloor$ and $b=i-aM$.
Two observations can be made from the structure of $\mathbf{B}_k$. First, $\mathrm{rank}\big(\mathbf{B}_k\big)\leq A_k$, which indicates that there are only $A_k$ columns in $\mathbf{B}_k$ are linearly independent. Second, each sub-matrix block $\hat{\mathbf{P}}^H_k\hat{\mathbf{B}}_i\hat{\mathbf{P}}_k$ is a $A_k$-dimension rank-one matrix since $\mathbf{u}_b$ and $\mathbf{H}_{i-M^2}$ are both rank-one. This reveals that linear independence cannot occur inside each sub-matrix block. Therefore, only $A_k$ sub-matrix blocks denoted by $\{\hat{\mathbf{P}}^H_k\tilde{\mathbf{B}}_j\hat{\mathbf{P}}_k\}$ with $j\in\mathcal{J}=\{1,\dots,A_k\}$ are linearly independent. The remaining $M^2+K-A_k$ sub-matrix blocks are linearly dependent, which can be expressed by the linear combination of $\{\hat{\mathbf{P}}^H_k\tilde{\mathbf{B}}_j\hat{\mathbf{P}}_k\}$ for $\forall j$. On this basis, $\mathrm{Tr}\big(\mathbf{B}_k\mathbf{X}_k\big)=0$ can be reformulated as $\mathrm{Tr}\big(\hat{\mathbf{P}}^H_k\tilde{\mathbf{B}}_j\hat{\mathbf{P}}_k\mathbf{X}_k\big)=0$ for $\forall j$. Furthermore, since the residual sub-matrix blocks are dependent, it follows that  
\begin{equation}
\mathrm{Tr}\big(\hat{\mathbf{P}}^H_k\hat{\mathbf{B}}_i\hat{\mathbf{P}}_k\mathbf{X}_k\big)=0, ~ 1\leq i\leq M^2+K.
\label{Zero-null}
\end{equation}

A semi-definite matrix  $\bar{\mathbf{W}}_k$ can be constructed as follows:
\begin{equation}
\bar{\mathbf{W}}_k = \hat{\mathbf{P}}_k\left(\mathbf{I}-\frac{1}{\delta_k}\mathbf{X}_k\right)\hat{\mathbf{P}}^H_k,
\label{rank-reduced}
\end{equation}
where $\delta_k$ is the maximal eigenvalue of $\mathbf{X}_k$ and $\mathrm{rank}\big(\bar{\mathbf{W}}_k\big)\leq A_k-1$.  $\bar{\mathbf{W}}_k$ is semi-definite due to $\mathbf{I}-\frac{1}{\delta_k}\mathbf{X}_k\succeq \mathbf{0}$. Combining equations (\ref{Zero-null}) and (\ref{rank-reduced}), one finds 
\begin{align}
\mathrm{Tr}\left(\hat{\mathbf{B}}_{i}\bar{\mathbf{W}}_k\right)&=\mathrm{Tr}\left(\hat{\mathbf{B}}_{i}\hat{\mathbf{P}}_k\hat{\mathbf{P}}^H_k\right)-\frac{1}{\delta_k}\mathrm{Tr}\left(\hat{\mathbf{P}}^H_k\hat{\mathbf{B}}_{i}\hat{\mathbf{P}}^H_k\mathbf{X}_k\right)\notag\\&\overset{(a)}{=}\mathrm{Tr}\left(\hat{\mathbf{B}}_{i}\hat{\mathbf{W}}_k\right).
\end{align}
This manifests that the rank-reduced matrix $\bar{\mathbf{W}}_k$ can meet constraints (\ref{ob_b3}), (\ref{ob_c3}), and (\ref{ob_e3}) in problem (\ref{linear_p3}). Meanwhile, the communication rate remains unchanged, so $\bar{\mathbf{W}}_k$ can achieve the same communication performance with $\hat{\mathbf{W}}_k$ but with a lower rank. Repeating the above procedure, the rank-one solution can be obtained. 

	\ifCLASSOPTIONcaptionsoff
	\newpage
	\fi
	
	\bibliographystyle{IEEEtran}
	\bibliography{references}
	
\end{document}